\documentclass[draft,12pt]{article}
\usepackage{psfig,amsfonts,amssymb,amsbsy}
\usepackage[mathscr]{eucal}

\def\thedraft{
  \PassOptionsToClass{draft,12pt}{article}
  \usepackage[nomarkers,noheads]{endfloat}
  
  \def\mybibsep{}
  \def\figplace{ht} 
  \def\figw{6in}
  
  \topmargin -0.5in
  \textheight 9in
  \textwidth 6.5in
  \evensidemargin 0in   
  \oddsidemargin 0in
  \def\abstractwidth{6.5in}
  \def\smallersize{\normalsize}
  \def\baselinestretch{1.66} 
}

\thedraft

\newcommand{\infinity}{\infty}
\newcommand{\LSF}{$\lambda$-structure factor}
\newcommand{\LB}{{\boldsymbol{\nabla}^2_{\!\mbox{\tiny{LB}}}}}

\newcommand{\Gkgst}{{G_{\!\mbox{\tiny $K_g$}}(s,t)}}

\newcommand{\Gktwost}{G_{\!\mbox{\tiny $K$}}^2(s,t)}
\newcommand{\Gktotst}{{G_{\!\mbox{\tiny $\vec K$}}(s,t)}}
\newcommand{\vv}{{\vec{\nu}}}
\newcommand{\nvv}{\nu} 
\newcommand{\X}{{\vec{X}}}
\newcommand{\vK}{{\vec{K}}}
\newcommand{\vR}{{\vec{R}}}
\newcommand{\vsR}{{\tilde{R}}}
\newcommand{\sqrtgmin}{\sqrt g_{\mbox{\scriptsize{min}}}}
\newcommand{\dtmax}{{\Delta t_{\mbox{\scriptsize{max}}}}}
\newcommand{\half}[1]{{\scriptstyle#1\frac{1}{2}}}

\newcommand{\tangent}{{\hat\tau}}
\newcommand{\normal}{{\hat n}}
\newcommand{\stangent}{{\tilde\tau}}
\newcommand{\snormal}{{\tilde n}}

\newcommand{\de}{\mbox{d}}
\newcommand{\ddu}[2]{{\partial #2\over \partial u^{#1}}}
\newcommand{\dRda}{{\partial \vR\over \partial \alpha}}
\newcommand{\ddt}[1]{{\left({\partial{#1}\over\partial t}\right)_{\!\!\alpha}}}
\newcommand{\ddtbig}[1]{{\left({\partial\over\partial t}{#1}\right)_{\!\!\alpha}}}
\newcommand{\dda}[1]{{\partial{#1} \over \partial\alpha}}
\newcommand{\ddabig}[1]{{{\partial\over\partial\alpha}{#1}}}
\newcommand{\dds}[1]{{\partial{#1} \over \partial s}}
\newcommand{\ddsbig}[1]{{{\partial\over\partial s}{#1}}}
\newcommand{\dddss}[1]{{\partial^2{#1} \over \partial s^2}}
\newcommand{\dddaa}[1]{{\partial^2{#1} \over \partial\alpha^2}}
\newcommand{\dddaabig}[1]{{{\partial^2 \over \partial\alpha^2}{#1}}}

\def\slabel{\label}
\newcommand{\beq}{\begin{equation}}
\newcommand{\eeq}{\end{equation}}
\newcommand{\beqa}{\begin{eqnarray}}
\newcommand{\eeqa}{\end{eqnarray}}
\newcommand{\noeq}[1]{(\ref{eq:#1})}
\newcommand{\eq}[1]{Eq.~(\ref{eq:#1})}
\newcommand{\Eq}[1]{Equation (\ref{eq:#1})}
\newcommand{\fig}[1]{Fig.~\ref{fig:#1}}
\newcommand{\Fig}[1]{Figure \ref{fig:#1}}
\newcommand{\eqprot}[1]{eq.~(\protect\ref{eq:#1})}
\newcommand{\dontput}[1]{}

\newcommand{\ifinteg}{{\sf IFinteg}}
\newcommand{\geninterf}{{\sf geninterf}}

\begin{document}

{\Large\bf Kinetics of phase ordering on curved surfaces}\\

{\bf Oliver Schoenborn and Rashmi C.\ Desai}
\footnote{Department of Physics, University of Toronto, Toronto, Ontario
          M5S 1A7 CANADA \\ email: desai@physics.utoronto.ca}\\
\begin{center}
\begin{minipage}{\abstractwidth}
  \smallersize
  \parindent 0pt
  {\em \today}
\vspace{2ex} \hrule \hrule \vspace{2ex}
An interface description and numerical simulations of model A kinetics are
used for the first time to investigate the intra-surface kinetics of phase
ordering on corrugated surfaces.  Geometrical dynamical equations are
derived for the domain interfaces.  The dynamics is shown to depend
strongly on the local Gaussian curvature of the surface, and can be
fundamentally different from that in flat systems:  dynamical scaling
breaks down despite the persistence of the dominant interfacial undulation
mode; growth laws are slower than $t^{1/2}$ and even logarithmic; a new
very-late-stage regime appears characterized by extremely slow interface
motion; finally, the zero-temperature fixed point no longer exists,
leading to metastable states.  Criteria for the existence of the latter
are derived and discussed in the context of more complex systems.
  \vspace{2ex} \hrule \hrule \vspace{2ex}
\end{minipage}
\end{center}

%
%

\section{Introduction}

Many two-dimensional surfaces exhibit internal degrees of freedom which
allow for phase ordering or phase separation to occur within the static or
dynamic surface.  Examples are lipid bilayer membranes\cite{lipowskyS95},
crystal growth on curved surfaces\cite{peczakGL93}, and thin film
deposition\cite{petroffM96}.  Interaction between the shape and internal
degrees of freedom of the surface are believed to play an important role in
such systems, initiating, modifying or eliminating chemical or physical
intra-membrane domain-ordering processes.  The dynamics can be induced not
only by varying the temperature, but also by changing such variables as the
pH or ionic concentration of an aqueous embedding solution (in the case of
lipid bilayers), or the lattice mismatch and deposition rate (film growth).
Experimentally\cite{gebhardtGS77}, both spinodal decomposition in binary
mixtures and phase ordering in one-component systems have been observed.
Phase ordering is more relevant to non-fluid phases and, in the case of
biological lipid membranes, is possibly involved in the control of enzyme
adsorption and protein-enzyme interaction on the surface of a bilayer.

For curved surfaces, the consequences of the interplay between surface shape
and intra-surface pattern formation are at present largely unexplored in
literature.  The small amount of experimental results are rudimentary, and
most of the theoretical work has been limited, due to the mathematical and
numerical challenges involved, to equilibrium models and shape-perturbation
calculations around equilibrium.  Many of the theories assume that the
domains are already formed (see, for instance, \cite{leiblerA87}).
Typically, some form of bilinear coupling between an intra-membrane order
parameter and the local geometry of the surface, such as mean curvature, is
used.  When kinetics are considered, one obtains a dynamical equation of
state similar to that of the Random-Field Ising model with long-range
correlations\cite{hobbie96}, but with the long-range correlations entering
through the gradients of the intra-surface
order-parameter\cite{schoenbornD97}.  One result has been the creation of
shape phase-diagrams\cite{lipowskyS95,gomperS94}.  Recently, some
researchers have done simulations to investigate shape change in surfaces
made of two types of lipids\cite{taniguchi96,kumarR98}.  These simulations,
relying on Monte Carlo and bulk Langevin equations, are extremely computer
intensive, limiting the possible complexity of the surfaces or the run time
of the simulations to experimentally irrelevant cases.

Any dynamics occurring in curved spaces invariably introduces new and
non-trivial concepts and effects which require considerable care and
exploration.  The problem of how the dynamical shape of a surface and an
intra-membrane ordering process may affect each other is complicated.  We
therefore first consider the simplest case where there is no explicit
coupling between the two, and new correlations arise through the geometry of
the curved surface.  Thus, rather than inquiring about the shape-change when
the membrane is inhomogeneous, we investigate the influence of non-euclidean
geometry on pattern formation and phase ordering kinetics when it occurs on
a static curved surface.  By comparing the results with known results for
model A kinetics within a flat (euclidean) surface, we identify the sole
effects arising from surface curvature.  We hope that this will subsequently
allow for a systematic extension of the problem in which the surface is not
static and the explicit coupling between the intra-membrane order parameter
and the local geometry of the surface is also included.

More explicitly, we report in this paper novel results of a ground-up study
of relaxational pattern-formation occurring within static two-dimensional
surfaces, using model A\cite{hohenbergH77} as a starting point, {\em
without} any explicit coupling to the local geometry other than through
diffusion\cite{schoenbornD97}.  In section 2, we setup the analytical
framework for the intra-surface phase ordering kinetics on curved surfaces.
The bulk kinetics are described by the Non Euclidean Model A 
equation, Eq.(1).  Systems with phase ordering instabilities, with
homogeneous initial state and small random fluctuations, often evolve to
form domains separated by sharp interfaces.  The bulk part of the domains
equilibrates rapidly and the interface width saturates within a very short
time.  Further time evolution involves only interface dynamics in which
interfacial widths hardly change, but the total interfacial length decreases
in order to minimise the system free energy.  In order to take advantage of
this type of kinetics, we deduce in section 2, a set of equations [Eqs.
(2), (13) and (28)], which are equivalent to the bulk Non Euclidean Model A
Eq.(1).  In this
interface description, local geodesic curvature of the interface, local
velocity of the interface and total interfacial length $L(t)$ play a central
role.  One of the novel quantities that we investigate in detail in this
paper (apparently for the first time in literature) is the Geodesic
Curvature Autocorrelation Function.  Equation (2) is the
generalization of the well-known Allen-Cahn equation for intra-surface
interface dynamics in curved surface systems.  Equations (13) and (28)
couple to Eq.(2) due to the geometry of interfaces on a curved surface.

In section 3, we explain how the numerical simulations were carried out:
the interface description became pivotal to the feasibility of simulations,
providing gains of 50 -- 100 in runtime with regards to the standard bulk
description, for the system sizes that were explored.  We also describe the
checks on the numerical algorithm that we performed for special cases where
analytic results are known.

For all the results described in sections 4 to 7, we first integrated the
bulk Non Euclidean Model A Eq.(1) for a short time to generate sharp 
interfaces on a variety
of surfaces, extracted the interfaces, and then evolved them in time using
three coupled interface equations.  Three important quantities used in these
sections are $\vK, \vK_g$, and $K_G$, which must not be confused:  the
intra-surface interfaces can be characterized {\em locally} by their total
interface curvature $\vK$ or by their geodesic curvature $\vK_g$, while a
curved surface, on which those interfaces move, can be described locally by
its Gauss curvature radius $K_G$\cite{defineKG}.  These simulation results
discuss how $K_G$ affects interface evolution and what are some of the new
features of Non Euclidean Model A dynamics on corrugated surfaces, 
such as non-power-law
domain growth, a new dynamical regime characterized by extremely slow
interface motion, breakdown of dynamical scaling, time-dependence of various
dynamical lengths not relevant in flat systems, and metastable interface
configurations and activated hopping.  Two measurements of interest for
characterizing the domain morphology are introduced, namely the
autocorrelation functions for $\vK_g$ and $\vK$.  We give some quantitative
and qualitative analytical explanations for these results while pointing to
some others which require further study.


\section{Interface Equations}

It was shown in \cite{schoenbornD97} that on curved surfaces, the equation
for model-A kinetics (sometimes refered to also as the time-dependent
Ginzburg-Landau equation, or TDGL for short) must be written as
\beq
  M^{-1}{\partial\phi\over\partial t} = \phi - \phi^3 + \xi^2\LB \phi
  \label{eq:NEMA}
\eeq
where $M$ and $\xi$ are positive phenomenological constants, $t$ is time and
$\phi$ is the intra-surface order parameter, while $\LB$ is the
Laplace-Beltrami operator\cite{laugwitz65}.  Some phenomenological
parameters have been eliminated by appropriately rescaling space, time, and
$\phi$\cite{guntonSS83}.  We refer to \eq{NEMA} as the Non-Euclidean Model A
equation.  The order parameter could be for instance the local
magnetization at the surface of a corrugated anisotropic Ising ferromagnet,
following a quench from a high-temperature, disordered (i.e.~paramagnetic)
state in thermal equilibrium, to a low-temperature, thermodynamically
unstable region of the temperature-magnetization phase-diagram..  In this
equation, there is no constraint on the average order parameter per unit
area as a function of time.  This differs from the well known model B which
describes spinodal decomposition in binary mixtures\cite{guntonSS83}, and
for which the order parameter is conserved.  From \eq{NEMA} we derived in
\cite{schoenbornD97} a Non-Euclidean Interface Velocity equation for
interfaces on curved surfaces,
\beq \vv = M \xi^2 \vK_g \label{eq:NEIV} \eeq
where $\vv$ is the local interface velocity and $\vK_g$ is the local {\em
geodesic} curvature of the interface on the surface.  \Eq{NEIV} reduces to
the well-known Allen-Cahn equation in the Euclidean limit\cite{allenC79}.
Note that the derivation of \eq{NEIV} given in \cite{schoenbornD97},
although concise, is not as geometrically transparent as the different one
given in \cite{schoenborn98}.

Further insight into the interface dynamics can be gained by considering the
evolution of $\vK_g$ in time.  We give here a parameterization-invariant
derivation of geometrical dynamical equations valid for any model-A
interface on any Riemannian surface.  To do this, it is necessary to
parameterize the interface.

Because, as indicated by \eq{NEIV}, the interface moves normal to itself, it
is convenient to use a parameterization of interfaces which exploits this
feature:  the normal gauge\cite{browerKKL84}.  This dynamical constraint
defines the gauge, whose units of length are therefore not physical units,
and change both in time and space.  We denote the parameter of the normal
gauge by $\alpha$.  Physical measurements, on the other hand, are always
done using real (hence constant) units of distance.  The parameterization
which satisfies this is the arclength gauge, whose parameter we denote by
$s$.

The two gauges are linked by a {\em metric} which we denote $\sqrt g$.
Denote the curvilinear coordinate system on the surface by
$(u^1,u^2)\equiv(u,v)$.  The surface is described by a series of
three-dimensional vectors $\X = [x(u,v),y(u,v),z(u,v)]$, while the interface
is described by a series of two-dimensional ``vectors'' $\vsR(\alpha) \equiv
[u(\alpha), v(\alpha)]$ or equivalently by a series of three-dimensional
vectors $\vR(\alpha) \equiv [x(\alpha), y(\alpha), z(\alpha)]$.  We use the
convention of an arrow to denote a three-dimensional vector, and a tilde a
two-dimensional vector in the tangent space of the surface.
\begin{figure}[\figplace]
  \centerline{\psfig{figure=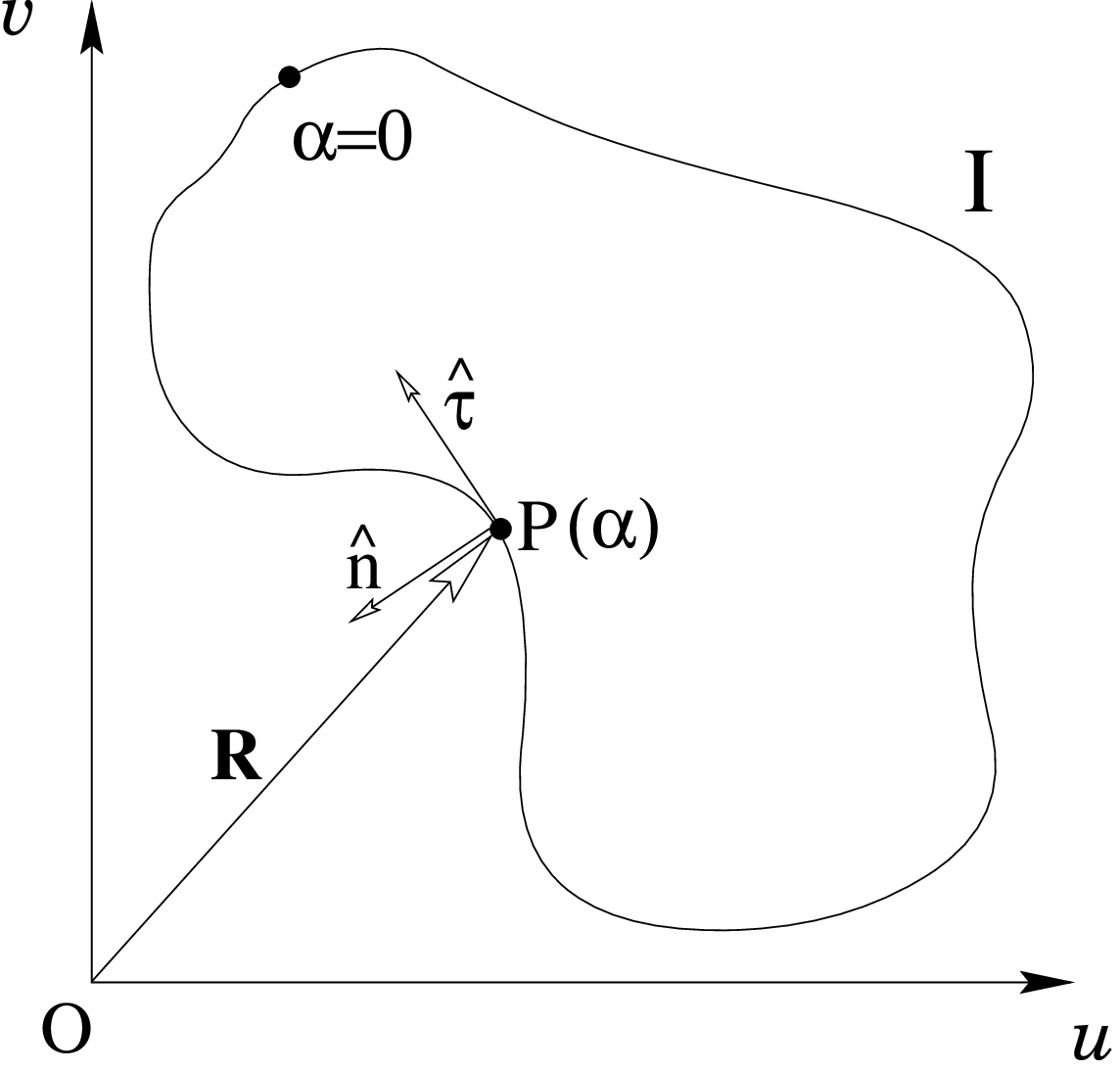,width=\figw,angle=270}}
  \caption{An interface $I$ in the $(u,v)$ space of the surface.  A point
	   $P$ of the interface has parameter $\alpha$ and is given by a
	   vector $\vR(P)$ (bold $R$ in figure).  $\tangent$ and $\normal$
	   are the tangent and normal vectors to $I$ at $P$.}
  \label{fig:interf}
\end{figure}
A typical model-A interface, labelled $I$, is shown in \fig{interf}, along
with the vector $\vR$ going from some arbitrary origin on the surface to a
point $P(\alpha)$ on $I$, and the tangent $\tangent$ and normal $\normal$ to
the interface at $P$.  Every point of an interface corresponds to a unique
$\alpha$ which does not change in time, since the interface moves normal to
itself.

The relationship between the differential elements of length $\de s$ and
$\de\alpha$ is given by
\beq
  \de s = \sqrt g \;\de\alpha.
  \label{eq:dsvsda}
\eeq
where the metric is defined by
\beq
  g(\alpha,t) \equiv \left| \dda{\vR} \right|^2.
  \label{eq:metricinterf}
\eeq
Because $\de\alpha$ is constant in time while $\de s$ is not, $g$ changes in
time.  Its evolution equation must be determined, making use of the
requirement that a point of coordinate $\alpha$ moves only perpendicularly
to the interface.  This can be done in the following way.

First note that $\tangent$ and $\vv$ are defined as:
\beqa
  \tangent & \equiv & \dds\vR = (u', v') \slabel{eq:tau_vs_R} \\
  \vv      & \equiv & \ddt\vR            \slabel{eq:v_vs_R},
\eeqa
where a prime denotes differentiation relative to the arclength $s$.  Taking
the time derivative at constant $\alpha$ on both sides of \eq{metricinterf},
we have
\beqa
  \ddt g & = &  2 \dRda \cdot \ddtbig{\left(\dRda\right)}\nonumber \\
   & = & -2 \dddaa{\vR} \cdot \vv.
  \label{eq:dgdt1}
\eeqa
Now since
\beqa
  \alpha'    & \equiv & \dds  \alpha = {{1}\over{\sqrt g}}  \\
  \alpha'{}' & \equiv & \dddss\alpha = {-1\over 2g^2}\dda{g},
\eeqa
the Laplacian in arclength $s$ is 
\beq
  \dddss{} = {1\over g}\dddaa{} + \alpha'{}'\dda{}
  \label{eq:d2svsd2a}
\eeq
so that from the definition of the (three-dimensional) interface 
curvature vector $\vK \equiv \dddss\vR$,
\beq
  \dddaa\vR = g \left( \vK - \tangent \alpha'{}' \sqrt g \right).
  \label{eq:d2Rda2}
\eeq
Substituting \eq{d2Rda2} in \eq{dgdt1}, using the orthogonality between
$\vv$ and $\tangent$ and rewriting \eq{dgdt1} in terms of $\sqrt g$ rather
than $g$ one obtains
\beq
  \ddt{\sqrt g} = -\sqrt g\vK \cdot \vv,
  \label{eq:dgdt2}
\eeq
which further reduces to 
\beq
  \ddt{\sqrt g} = -\sqrt g\vK_g \cdot \vv,
  \label{eq:dgdt3}
\eeq
when one notes that $\vv$ is parallel to $\normal$ and lies within the
surface, whereas $\vK$ has a component parallel to $\vv$ {\em and} one
normal to the surface.  I.e., if the normal to the surface is denoted by
$\hat N$, then $\vK = |\vK_g|\normal + |\vK_n|\hat N$, with $\vK_n$ the
three-dimensional curvature of a geodesic tangent to $I$ at $P$.  \Eq{NEIV}
and the evolution equation for $g$, \eq{dgdt3}, are two of the three
interface equations that are used in our interface description.

Now we proceed with a similar but lengthier derivation for $(\partial K_g /
\partial t)_\alpha$.  The dot product of $\normal$ with \eq{d2Rda2} yields
\beq
  {\normal\over g} \cdot \dddaa\vR = |\vK_g|.
  \label{eq:gK}
\eeq
Applying the time derivative and using the chain rule,
\beq
  \ddt{|\vK_g|} =  {\normal\over g} \cdot\dddaabig{\ddt\vR} 
		 - {\normal\over g^2}\cdot\dddaa\vR \ddt g 
		 + {1\over g}\ddt\normal\cdot\dddaa\vR .
\eeq
The second term is simplified via \eq{gK} and \eq{dgdt3}.  The third one is
simplified by using \eq{d2Rda2} and noting that $\normal$ can only change in
direction, hence $\ddt\normal$ is oriented along $\tangent$.  With those
simplifications one gets
\beq
  \ddt{|\vK_g|} =  {\normal\over g} \cdot\dddaa\vv + 2|\vK_g|\vK_g\cdot\vv
	   - {\sqrt g}\alpha'{}' \tangent \cdot \ddt\normal
  \label{eq:dKdt0}
\eeq
The second derivative of $\vv$ is
\beq
  \dddaa\vv \equiv \dddaabig{(\nvv \normal)} = 
  \dddaa\normal \nvv + 2\dda\normal \dda \nvv + \normal \dddaa \nvv.
  \label{eq:ddvdaa}
\eeq
The first term is simplified by writing the second derivative of $\normal$
in terms of {\em arclength} via \eq{dsvsda}, applying the chain rule and
using the two Frenet equations for curves in Riemannian
spaces\cite{laugwitz65}
\beqa
  \dds \tangent & = &  |\vK_g| \normal = \vK_g \\
  \dds \normal  & = & -|\vK_g| \tangent.
  \label{eq:NEFrenetTau}
\eeqa
This yields
\beq
  \dddaa\normal = -\sqrt g \tangent \ddsbig{(|\vK_g|\sqrt g)} - g |\vK_g| \vK.
  \label{eq:ddndaa}
\eeq
The second term of \eq{ddvdaa} vanishes when dot-multiplied with $\normal$
given the second Frenet equation.  After substitution of \eq{ddvdaa} and
\noeq{ddndaa} in \noeq{dKdt0}, we have
\beq
  \ddt{|\vK_g|} = {1\over g} \dddaa \nvv + |\vK_g| \vK_g\cdot\vv 
		 - \sqrt g \alpha'{}' \tangent \cdot \ddt\normal.
  \label{eq:dKdt1}
\eeq

Finally we show that $\sqrt g \tangent \cdot \ddt\normal$ in the last term
of \eq{dKdt1} satisfies
\beqa
  \sqrt g \tangent \cdot \ddt\normal
    & = & - \sqrt g \dds \nvv\\
    & = & - \dda \nvv
  \label{eq:dndtfinal}
\eeqa
It is easiest to show this by first deriving $\dds\nvv$.  The norm of the
velocity is
\[
  \nvv = \normal\cdot \ddt\vR
\]
Taking the arclength derivative and using the chain rule, 
\beqa
  \dds \nvv  
    & = & {\normal\over\sqrt g} \cdot \ddabig{\ddt\vR}+\ddt\vR\cdot\dds\normal\\
    & = & {\normal\over\sqrt g} \cdot \ddtbig{\left(\dda\vR\right)}
\eeqa
since $\vv$ is orthogonal to $\dds\normal$.  Using $\tangent=\dds{\vR}$ and
the chain rule again yields
\beqa
  \dds \nvv  
    & = & \normal\cdot\ddt\tangent\\
    & = & -\tangent\cdot\ddt\normal
\eeqa
where the second equality is obtained by using
$\de(\tangent\cdot\normal)/\de t=0$, completing the demonstration.

Thus, substituting \eq{dndtfinal} into \eq{dKdt1} and using \eq{d2svsd2a},
we get the last of the three curvature equations:
\beq
  \ddt{|\vK_g|}  =  \dddss \nvv + |\vK_g|\vK_g\cdot\vv 
  \label{eq:dKdtfinal}
\eeq
The three equations, \eq{NEIV}, \eq{dgdt3} and \eq{dKdtfinal}, are a coupled
set of curvature equations describing the evolution of interfaces in the
Non Euclidean Model A.  Note that in these equations, if two-dimensional 
vectors are used,
the scalar product must be properly defined via the surface's covariant
metric tensor $g_{ij} \equiv \ddu{i}{\X}\cdot\ddu{j}{\X}$, i.e.~$\tilde a
\cdot \tilde b \equiv a_ig^{ij}b_j = a^ig_{ij}b^j$, where an implicit sum
over alternate repeated indices is assumed.  \Eq{dKdtfinal} and \eq{dgdt3}
are purely geometric but only valid in the normal gauge and when $\vv$ is
parallel to $\normal$.

The first term on the right hand side of \eq{dKdtfinal} is diffusive 
and causes all
modes of $K_g$ to dissipate, except the zeroth mode ($K_g=0$) which is not
affected by it.  The diffusive term thus seeks to shrink interfaces either
into straight lines (on a curved surface, geodesics) which have zero $K_g$,
or perfect (i.e.~geodesic) circles which have constant $K_g$.  On the other
hand, due to the Non-Euclidean Interface Velocity equation, 
the second term on the right hand side
of \eq{dKdtfinal}
is cubic in $K_g$.  It seeks to increase the curvature and dominates when
the diffusive term is negligible, i.e.~for straight lines and circular
domains.  Hence it causes circular domains to shrink in radius, as is indeed
the case in Euclidean model A.  If the circular domain is also a geodesic,
so that $K_g=0$ everywhere along the interface, the domain will not change
in shape.  When the model-A dynamics proceeds from a quench of a system from
a high temperature, such that the domains are initially very disordered, the
diffusive term of \eq{dKdtfinal} dominates, broadly speaking, during the
early stages of interface motion, whereas the cubic term dominates only in
interfaces which have become circular.  However, this was verified in
numerical simulations to be only approximately true.

Another interesting aspect of the geometric equations is that substitution
of \eq{NEIV} into \eq{dgdt3} shows how $\ddt{\sqrt g}$ is always negative.
Therefore the distance between close-by interface points always decreases in
time, and by extension also the total length of an interface, which is given
by $L=\int\!\sqrt g \de\alpha$:
\beq
  \ddt{L} = -M\xi^2\int\!\! |\vK_g|^2\;\de s.
  \label{eq:dLdt}
\eeq
Without these results it is difficult to ascertain whether \eq{NEIV} is
consistent with the physical nature of the bulk equation for model A
kinetics, where the quantity of order-parameter gradients (equivalent to the
total length of interface in the system) should monotonically decrease in
time.

The apparent simplicity of the Non-Euclidean Interface Velocity equation 
can be misleading.  The
geodesic curvature of a line on a curved manifold depends both on the
position of that line and its orientation on the surface.  Therefore, $K_g$
introduces into the dynamics a new, position- and orientation-dependent
length scale.  This strongly suggests that Non Euclidean Model A dynamics 
cannot be
self-similar as it is in flat systems, and therefore that dynamical scaling
will not be observed on curved surfaces, except perhaps on self-affine
surfaces, such as found over a certain range of length scales in lipid
bilayer membranes.  Different surfaces should show different phase-ordering
kinetics.  This is further evidenced in \eq{dLdt} where $L$ --- proportional
to the reference length scale of the dynamics, if it is
present\cite{schoenbornD98a} --- depends on $\int |K_g|^2 \de s$.

Nonetheless, a study of Non Euclidean Model A on the torus manifold, 
reported in
\cite{schoenbornD97}, revealed no clear signature of the non-Euclidean
nature of the surface when investigated through the evolution of the
interface density (quantity of interface per unit area) and the 
$\lambda$-structure factor \cite{lambdasf} of the order parameter.  The
latter is a two-dimensional order parameter structure factor which
depends only on the intrinsic geometry of the surface and the
order-parameter configurations.  This absence of signature may be due not
only to the coarse-grained nature of the \LSF, but also to finite size
effects, since domains feel the geometry only when their size becomes
comparable to the torus size.  This suggests that surfaces with geometrical
features closer to those of model-A domains at early times should be
investigated.  Indeed, as we show in section \ref{sec:metastable},
interfaces can get caught around certain surface bumps, causing a drastic
slowing down and even immobilization of the model-A dynamics.  Before
investigating this, we outline the numerical methods used in the
simulations.

\section{Numerical method}

\dontput{
  In the normal gauge, the number of points which constitutes an interface
  is constant in time, but since interfaces are decreasing in length as a
  function of time, the distance between any two neighboring points also
  decreases in time, as shown shortly.  In the arclength gauge, the number
  of points which constitutes an interface decreases in time, since
  neighboring pairs of points are always separated by the same distance,
  meaning in this gauge the points have both a normal {\em and} tangential
  velocity component.  The resulting arclength gauge is always the one used
  to express physical results, though all manipulations are done via the
  normal gauge.
}

It is customary to simulate equation for model A kinetics by discretizing
them and evolving the order-parameter configurations via an Euler
integration scheme.  The main limitation of such discretization schemes is
that the time step for evolving the system is limited by the smallest space
mesh in the system.  On curved surfaces, this can be --- and most often is
--- a severe impediment, as the surface mesh can rarely be optimized so as
to be homogeneous.  Doing this involves computational challenges in its own
right.  Slightly more sophisticated methods such as spectral,
predictor-corrector and implicit schemes are possible but the stiffness of
the model-A kinetic equation still induces strong dependencies on spatial
mesh which make them unappealing.

We have found that using \eq{NEIV} as the basis for an interface description
has many advantages over a bulk description as given by a model A equation.
Namely, the quantity of information to manipulate at every time step is an
order of magnitude smaller at the beginning of the simulation than with a
bulk description, and rapidly decreases in time, while for the bulk
description it remains constant in time, even if the order-parameter
configuration has become completely homogeneous.  Secondly, the surface can
be discretized independently from the interfaces so that numerical
instabilities associated with the surface mesh disappear.  Thirdly, the
interfaces being one-dimensional, the integration algorithms remain fairly
simple and more sophisticated algorithms are much easier to implement.
Finally, certain quantities of interest, which we introduce below, are far
easier and faster to compute from the interfaces than from the bulk
configurations.

One inevitable drawback of an interface description is that it cannot
describe how interfaces form from the small order-parameter fluctuations
present just after the quench.  Therefore, the initial interface
configurations must be obtained independently.  The easiest way to do this
is of course to use the bulk description to evolve the order parameter for
only the relatively short time necessary for interfaces to become
established, and then extract the interfaces from the resulting
configurations.  Even for short integration times, this suffers from the
drawbacks mentioned above.  We give one alternative, in appendix C of
\cite{schoenborn98}, devoid of any bulk steps and dependent only on the
statistical properties of interfaces.

For this reason, initial interface configurations are always obtained by
integrating the bulk equation \noeq{NEMA} from $t=0^+$ to $t=17$, when the
domains have fully formed and interfaces can be extracted.  For integration
of bulk equations on curved surfaces, \cite{thompsonWM85} has proved
invaluable.  Once the bulk configurations are obtained, the interface
extraction is done with a program which we refer to as \geninterf.  A
full-length run, done by a program we refer to as \ifinteg, then takes the
interface configurations from $t=17$ to $t=10000$.  Such long runs were
necessary in Non Euclidean Model A due to the slowing down of the dynamics 
at later times.  A
typical initial interface configuration is shown in \fig{typical}, as would
be seen in flat systems and in surfaces whose geometric features are much
larger than the size of the initial interface convolutions.
\begin{figure}[\figplace]
  \centerline{\psfig{figure=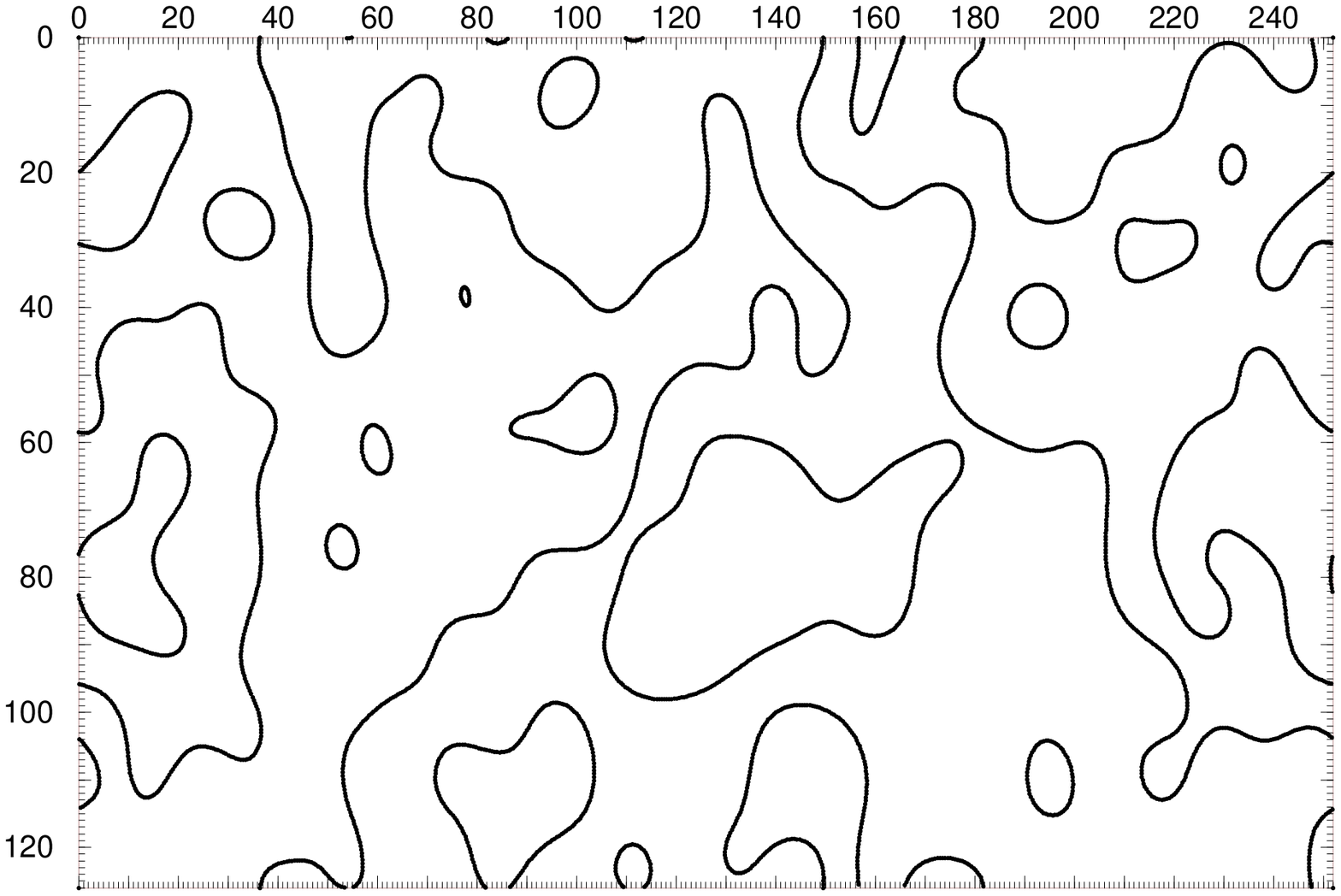,width=\figw,angle=180}}
  \caption{A typical early-time configuration of interfaces in a flat system.}
  \label{fig:typical}
\end{figure}

An interface represented by its ordered set of coordinate points
$\vsR(\alpha)$ is evolved in the tangent space $(u^1,u^2) (\equiv(u,v))$ of
the surface via the Non-Euclidean Interface Velocity equation by writing 
the latter in the form
\beq 
  \ddt{u^i} = M \xi^2 K_g n^i
  \label{eq:NEIVK} 
\eeq
where $n^i$ is the $i^{th}$ component of the interface normal $\snormal$,
and $K_g$ is given, in compact notation, by
\beq
  K_g = \epsilon_{ij} u^i{}'(u^j{}'{}' + \Gamma^j_{kl} u^k{}'u^l{}')
  \label{eq:kg}
\eeq
where again implicit summation over repeated indices $\{i,j=1,2\}$ is
assumed.  The prime denotes differentiation with respect to arclength.
Arclength is always the physical length of the curve as measured in the
Euclidean embedding space of the surface.  $\epsilon_{ij}$ is $(j-i)\sqrt
{g_m}$, with ${g_m}=g_{11}g_{22}-g_{12}g_{12}$ the determinant of the metric
of the {\em manifold} rather than that of the interface, which is $\sqrt g$.
\Eq{kg} defines $K_g$ as a {\em signed} scalar, therefore $\snormal$ is
defined as $\stangent$ rotated by $\pi/2$ counterclockwise and requiring
$\snormal\cdot\stangent=0$, leading to
\beq
  \normal = {1\over\sqrt{g_m}}(-v'g_{22}-u'g_{12}, u'g_{11}+v'g_{12}).
  \label{eq:NEn}
\eeq

The numerical integration of \eq{NEIVK} for the $j^{\mbox{\scriptsize th}}$
interface point is done by first computing $\sqrt{g_m}$ and $\sqrt g$ at the
point, using these to obtain $K_g$ from \eq{kg} and $\snormal$ from \eq{NEn},
and then by moving the point via the time map
\beq
  u^i(j,t+\Delta t) = u^i(j,t) + \Delta t M\xi^2 K_g(j,t) n^i(j,t); i=1,2
\eeq
This Euler scheme becomes unstable if the discrete time step $\Delta t$ is
too large.  With the $\alpha$ parameter equal to the index of a point along
the interface, $\sqrt g$ is the physical distance between points.
Therefore, each interface is evolved with its own time step $\Delta t =
0.4\dtmax$, with
\beq
  \dtmax \equiv \half{}(\sqrtgmin)^2
\eeq
and $\sqrtgmin$ is the smallest distance between two neighboring points of
the discrete interface.  As the integration proceeds, $\sqrtgmin$ decreases,
which forces a decrease of $\Delta t$.  For this reason, the interface is
remeshed with all points separated by a distance of 1 unit whenever
$\sqrtgmin<0.5$.  The remeshing algorithm can be made very efficient, one
remeshing requiring barely more than a few integration steps.  In flat
systems, the interface description is roughly 5 times faster than the bulk
description.  On curved surfaces, the gain increases dramatically:  it was
50 or 100-fold in the simulations reported here, and will be much higher for
more complex surfaces.  Simulations using only the {\em bulk} description
require, for a batch of 40 runs on the type of surfaces used here, on the
order of 60 to 80 days on an HP9000s735.

The integration algorithm was verified for correctness and accuracy through
several independent checks.  Notably, the Non-Euclidean Interface Velocity
equation can be solved
exactly for a circular domain in a flat system.  The simulation of such a
domain yielded a curve indistinguishable from the theoretical prediction.
Statistical runs starting from random initial order-parameter configurations
were done with both bulk and interface descriptions and compared by
measuring the amount of interface per unit area and comparing interface
configurations and curvatures.  Differences smaller than the interface width
were found for the configurations, while the dynamical growth exponent of
domains in flat systems was found to be $0.48\pm0.01$ with the interface
description, closer to the theoretical prediction of 1/2 than the bulk
numerical integration result of $0.45\pm0.02$.  The interface description is
hence favored if only in accuracy.

Simulations were also done on the torus manifold with both the bulk and
interface descriptions and found to be once more identical apart from the
similar difference in the dynamical exponents.  In general, the interface
description provides more accurate results than the bulk description, as
seen by repeating the simulation of a band domain on a torus
manifold\cite{schoenbornD97,schoenborn98} for different surface mesh
coarseness.  The interface velocity from the bulk description converges
towards the Non-Euclidean Interface Velocity theoretical prediction 
as the surface mesh is refined,
while that from the interface description falls {\em exactly} on the
theoretical prediction for all surface meshes used.  This is shown in
\fig{veloc} where $\nu(t)$ is the interface velocity, and $\theta_I$ the
Interface position on the torus (the torus manifold with its coordinate
system and manifold are shown below in \fig{tor_param}).
\begin{figure}[\figplace]
\centerline{\psfig{figure=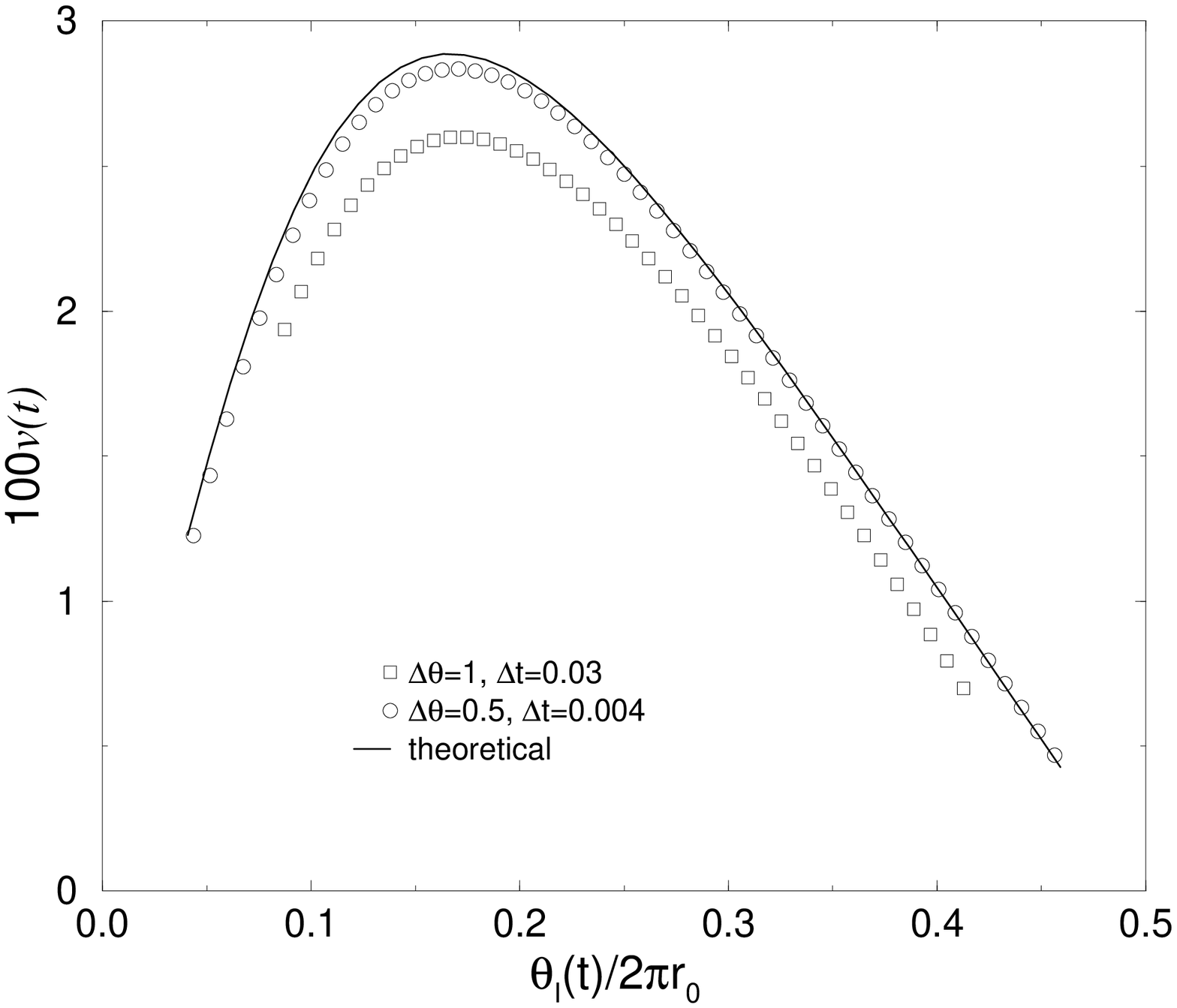,width=\figw,angle=270}}
  \caption{Interface velocity of stripe domain as a function of $\theta_I$
	   on the torus; $R=40, r_0=20$.  Numerical simulation with
	   $\Delta\theta=0.5$ (circles) and 1 (squares), and analytical
	   prediction (line) are shown.  At $t=0$ both top and bottom 
	   interfaces are near the outside equator of the torus.  }
  \label{fig:veloc}
\end{figure}

Finally, as a further accuracy check, a circular domain on a surface
consisting of one circular bump was simulated as the interface equations can
be solved analytically.  With the surface defined by
\beq
  \X = \left[u,v, A e^{-(u^2+v^2)/2\sigma^2}\right],
\eeq
with $A$ the amplitude of the bump and $\sigma$ its half-width, the
relationship between $K_g$ and the radius $R$ of the domain is
\begin{equation}
  K_gR = { 1\over
	    \sqrt{ 1 + 
	    \left({A\over\sigma}\right)^2 
	    \left({R\over\sigma}\right)^2 e^{R^2/\sigma^2}
	 }}
  \label{eq:kggauss}
\end{equation}
The comparison of the theoretical prediction and numerical simulation is
shown in \fig{gausstesta} where the product $K_gR$ is plotted {\em vs} $R$.
Two different surface meshes were used, showing what kind of surface
coarseness is sufficient for accurate results:  the longest surface tether
should be around 0.7 units long for best results, though 1.3 units gave
satisfactory results as well, as seen in the figure.
\begin{figure}[\figplace]
  \centerline{\psfig{figure=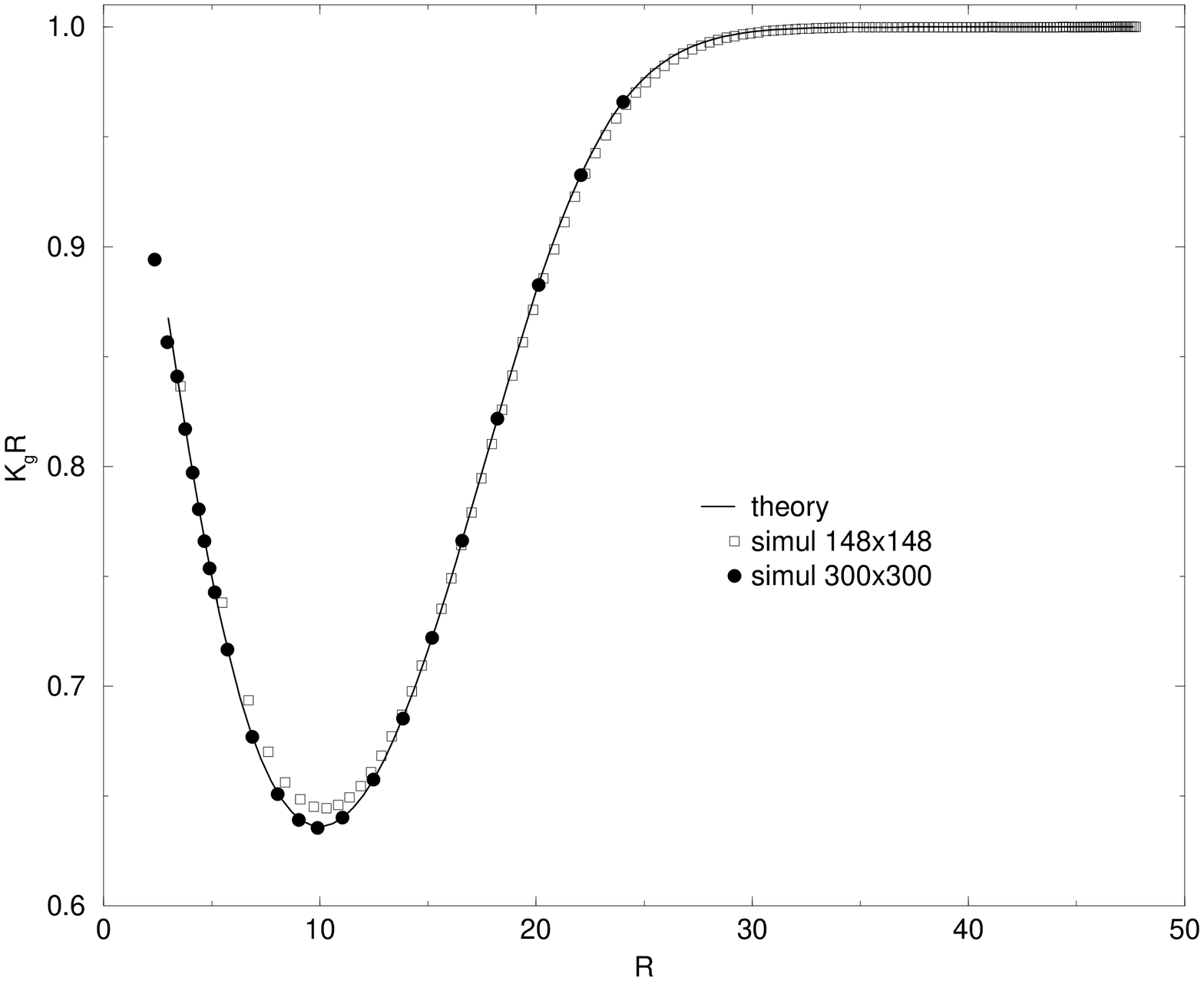,width=\figw,angle=270}}
  \caption{$K_gR$ of a circular domain as a function of its radius $R$, as
	   given by numerical integration on 2 different grid sizes; also
	   shown is analytical prediction \eqprot{kggauss}.  Error bars
	   too small to be seen.}
  \label{fig:gausstesta}
\end{figure}



\section{Effects of surface curvature on domain growth}

A naive interpretation of the Non-Euclidean Interface Velocity equation may 
suggest that Non Euclidean Model A dynamics
should always be slower than Euclidean dynamics:  $|\vK_g|$ is always less
than or equal to the total interface curvature $|\vK|$.  However diffusion
occurs faster (slower) in regions of negative (positive) surface Gaussian
curvature $K_G$\cite{defineKG}, because there is more (less) area available
for a given interface length in a region of negative (positive)
$K_G$\cite{david88}.  This suggests that interfaces arising from the 
Non Euclidean Model A 
equation should disappear more slowly where $K_G>0$, but {\em faster} where
$K_G<0$.  Yet, simulations of the Non Euclidean Model A on the torus manifold, 
starting from
random initial order-parameter configurations, showed no dependence
whatsoever on Gaussian curvature\cite{schoenbornD97}.  This may have been
due to the use of the $\lambda$-structure factor \cite{lambdasf}
as the measure of the two-dimensional Order Parameter Structure Factor
or to the combination of the topology of the torus with the
percolating domains.

To settle the matter, {\em one} ovoid interface was therefore simulated in
the $K_G>0$ region of the torus manifold, another in the $K_G<0$ region and
again in a flat system ($K_G=0$).  The torus manifold is convenient for many
reasons discussed in \cite{schoenbornD97}, most importantly that two
well-separated regions of oppositely signed Gaussian curvature can be
defined on the torus.  A drawing of part of a torus surface with coordinates
and parameters is shown in \fig{tor_param}.
\begin{figure}[\figplace]
  \centerline{\psfig{figure=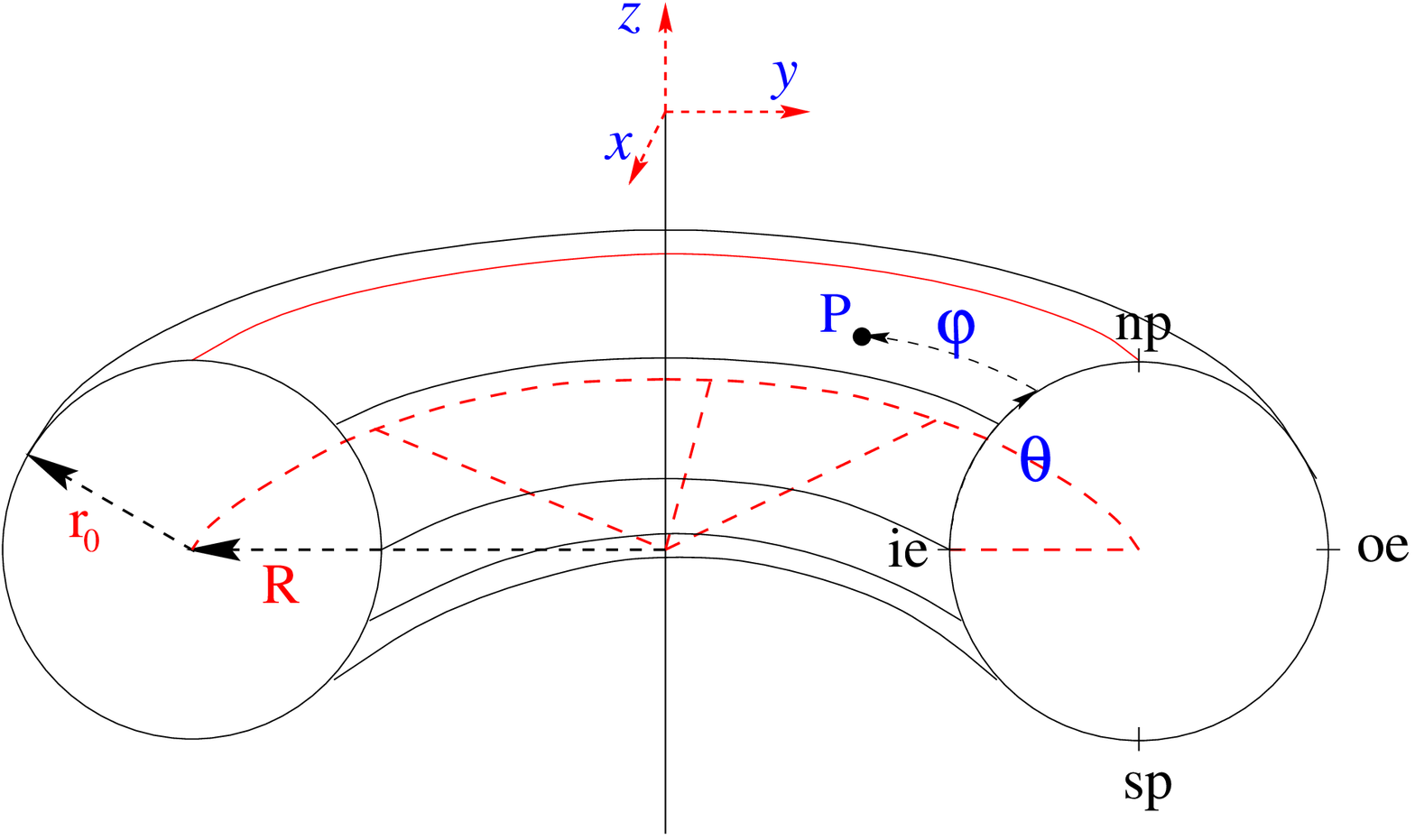,width=\figw,angle=270}}
  \caption{Torus manifold of long radius $R$ and small radius $r_0$.
	   Point $P$ on manifold has coordinates $(u^1,u^2) \equiv
	   (\theta,\varphi)$.  $\theta=0$ and 1 are at inner equator
	   (labelled ``ie''), $\theta=\pi r_0$ is at the outer equator
	   (labelled ``oe'').}
  \label{fig:tor_param}
\end{figure}
The small radius of the torus is denoted $r_0$ and the large one $R$.  The
initial radius of the ovoid is such that it is completely included in the
appropriate region of the torus, i.e.~$\pi r_0/2$ as measured in
$(\theta,\varphi)$ space (see \fig{tor_param}).  

\Fig{dLdtline} shows a log-log plot of $-\de L/\de t$ as a function of
$1/L$, where $L(t)$ is the time-dependent length of the ovoid interface in
the system.
\begin{figure}[\figplace]
  \centerline{\psfig{figure=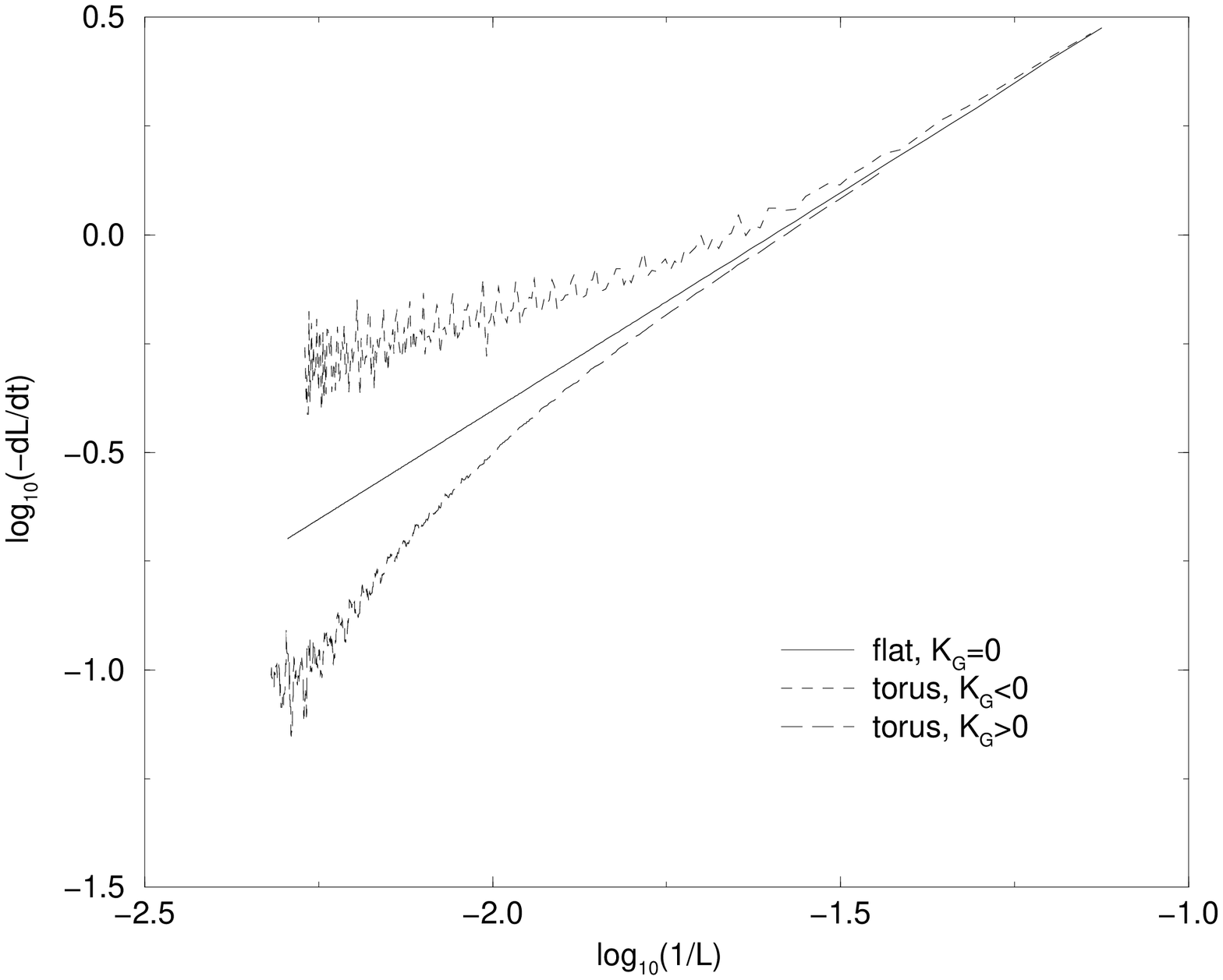,width=\figw,angle=270}}
  \caption{Log-log plot of $-\de L/\de t$ {\em vs} $1/L$.  Straight 
	   line is for simulation of flat system.  The scatter for nonzero
	   $K_G$ is due to numerical discretization of the surface.}
  \label{fig:dLdtline}
\end{figure}
A straight line is obtained in the flat system, as expected for Euclidean
model A.  However, differences as large as a factor of two can be seen at
early times with the curved regions, when the domains are large (roughly
20\% of system area).  The difference remains substantial until the domains
are very small.  
This suggests that the absence of visible non-Euclidean
effects reported in \cite{schoenbornD97} could be a characteristic of the
\LSF.  Note that the oscillations for the $K_G\neq0$ curves are due to the 
numerical discretization of the torus manifold.

The above simulation result can be understood analytically with the help of
\eq{dgdt3} by considering surfaces of {\em constant} Gaussian curvature
(i.e., a sphere, a flat plane, and a hyperbolic plane).  In this case, a
circular domain remains circular, as indicated by \eq{dKdtfinal}, i.e.~the
geodesic curvature is constant everywhere along the interface at any given
time (but changes in time).  Integrating both sides of \eq{dgdt3} along the
interface then yields $-\de L/\de t \sim LK_g^2$, where $K_g$ is a function
of time (note also \eq{dLdt}).  It is useful to express $K_g$ in terms of
$K_G$.  In a flat system $K_g^2=4\pi^2/L^2$ when the domain boundary has
length $L$.  For small $K_G\neq0$, dimensional analysis indicates the first
order correction to $K_g^2$ when in a curved surface should be of the form
$\gamma K_G$, with $\gamma$ a dimensionless geometrical factor of the order
of 1.  Moreover, when $K_G$ is increased from 0 (i.e.~the surface is a
sphere of decreasing radius) while keeping $L$ constant, the geodesic
curvature monotonically decreases, until it eventually becomes 0 when
$K_G=(2\pi/L)^2$.  On the other hand, when $K_G$ is increased negatively
from 0 (i.e.~the surface is a hyperbolic plane), the area available to the
circular domain increases, such that in order to keep $L$ constant, the
radius of the domain must be decreased, increasing the geodesic curvature.
This leads to the first order approximation for the time change of the
circular domain's interface length,
\beq
  -{\de L\over\de t} \sim 
    {4\pi^2\over L} \left(1 - {\gamma L^2K_G \over 4\pi^2}\right),
  \label{eq:dLdt_correction}
\eeq
with $\gamma>0$, providing a qualitative explanation the difference in
growth laws for the ovoid domains in regions of different Gaussian
curvature.

Given the physical origin of the influence of the surface Gauss curvature
$K_G$ on domain growth in the Non Euclidean Model A, 
much the same can be expected in
non-Euclidean model B as well as in other systems where diffusion and
interfaces are present on curved surfaces.  This slowing down for positive
$K_G$ was not observed in numerical simulations of pure model B on the
static sphere\cite{taniguchi96}, because of the short run times used.
However, similar (though not identical) Gauss curvature effects were seen in
simulations of crystal growth on toroidal geometries\cite{peczakGL93}.  In
the context of phase-ordering kinetics and phase-separation in lipid
bilayers, where diffusion is known to play a very important role
biologically, this dependence on $K_G$ could be of use in controlling the
rate of such processes as protein diffusion via membrane shape change, or
inducing enzyme-protein interaction in limited parts of a cell.


\section{Simulations on sinusoid surfaces}

When the initial order-parameter configuration is one of complete disorder,
the non-linear regime of Euclidean model A is characterized by the
relatively slow motion of sharp, convoluted interfaces delimiting domains of
$\phi=\pm1$ phases.  There are two characteristics of {\em
Euclidean}-model-A dynamics which are particularly important here.  First is
the self-similarity of the dynamics, leading to dynamical scaling:  system
configurations at a time $t_1$ look statistically identical to
configurations at an earlier time $t_0$, if they are rescaled lengthwise by
an appropriate factor.  Hence all dynamical lengths in the system have the
same time dependence, so that they can all be expressed in terms of one
arbitrarily chosen reference length scale $L(t)$.  The numerical value of
$L$ is not as important as its time dependence, which is the second
characteristic of relevance:  $L(t)\sim t^{1/2}$.

It is common to refer to $L$ as a dominant or typical length scale in the
dynamics, but the order parameter structure factor for Euclidean-model-A 
systems shows a peak at zero
wavenumber, indicating model-A dynamics does {\em not} have a {\em dominant}
length scale, only a unique {\em time dependence} for all dynamical lengths.
We have shown for the first time\cite{schoenborn98,schoenbornD98a}, by
considering the curvature correlations along the interface, that
Euclidean-model-A systems exhibit a dominant undulation mode not present in
the order-parameter structure factor.  If one is to talk of a dominant
dynamical length in model A, it is the wavelength of this undulation, not
the width of the order parameter structure factor.  
This suggests it has a different nature from that of
Euclidean model B, where the dynamical length is visible in the 
order parameter structure factor.
Notably, the absence of the dominant dynamical length in model A's 
order parameter structure factor may
be due to the absence of any phase information in it.

Now consider a surface consisting of a large array of bumps:
\beq
  \X = \left[ x, y, A\sin\left({2\pi x\over\lambda}\right)
	     \sin\left({2\pi y\over\lambda}\right) \right].
  \label{eq:sinusoidk20}
\eeq
Experimentally, fluid as well as crystalline lipid bilayers are known to
adopt a similar shape (called ``egg carton'') under certain
conditions\cite{goetzH96}.  Also, more complex surfaces such as self-affine 
surfaces common to fluctuating lipid bilayer membranes share many of the
qualitative features of sinusoid surfaces.

Simulations of Non Euclidean Model A on sinusoid surfaces, 
starting with random initial
order-parameter configurations, were done for several values of $A$ and
$\lambda$, but here we focus on the $\lambda=20, A/\lambda=0.2$ surface,
with system sizes of $100\times100$ to $300\times300$.

The normalized geodesic curvature autocorrelation function is 
defined by
\beq
  \Gkgst = {\langle K_g(s_0,t)K_g(s_0+s,t)\rangle 
		  \over
	    \langle K_g(s_0,t)^2\rangle},
\eeq
where $\langle\ldots\rangle$ denotes an ensemble average with an average
over all interface points $s_0$.  It is plotted in \fig{Gk2stRescL}, 
\begin{figure}[\figplace]
  \centerline{\psfig{figure=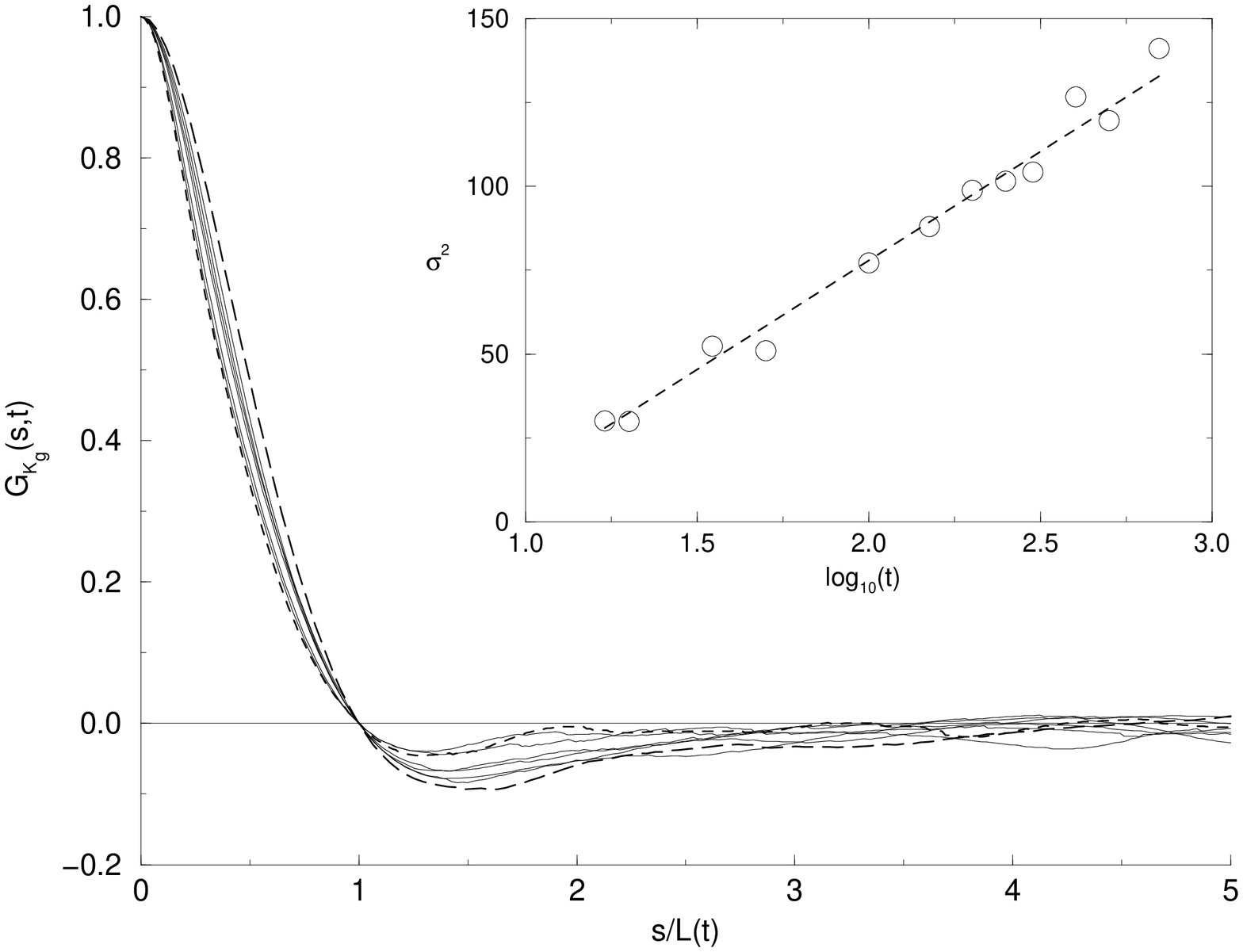,width=\figw,angle=270}}
  \caption{$\Gkgst$ at various times from $t=35$ (long dashed) to $t=400$
	   (short dashed), {\em vs} $s/L(t)$.  Sinusoid surface used is
	   \eqprot{sinusoidk20} with $A=4, \lambda=20$.  Inset:  Variance
	   of Gaussian part of $\Gkgst$ {\em vs} time (times between 17 and
	   700).}
  \label{fig:Gk2stRescL}
\end{figure}
where the horizontal axis has been rescaled with $L(t)$, defined here as
first zero of the geodesic curvature autocorrelation function, 
and the error bars (not shown for
clarity\cite{errbars}) are much smaller than the vertical offsets of the
curves.  In {\em flat} systems, as explained in
\cite{schoenborn98,schoenbornD98a}, all such geodesic curvature 
autocorrelation functions fall on top of one
another, due to dynamical scaling.  \Fig{Gk2stRescL} thus shows that for
the non-Euclidean case, dynamical scaling breaks down, as expected from our
discussion of \eq{NEIV} but contrary to runs on the torus manifold.  The
other important feature is that as time increases, the dominant interface
undulation mode monotonically decreases in intensity, signifying the
system's degree of order decreases as the interfaces explore ever larger
length scales.  This is sensible given that when the dominant length is much
smaller than the geometrical features of the surface, the interface
correlations decay fast enough to become negligible.  As the undulation
wavelength increases, the correlations become negligible at increasing
distances, introducing geometrical variations in the interface shape, thus
increasing the degree of disorder in the system.

These two features --- breakdown of dynamical scaling and existence of a
dominant dynamical length --- further confirm that the presence of a
dominant dynamical length does not guaranty dynamical scaling.

\Fig{Gk2stRescL} also indicates that the first zero of the geodesic
curvature autocorrelation function can no
longer be used as a reference dynamical length.  Two well-defined dynamical
lengths valid for interfaces on curved surfaces are the width $\sigma$ of
the Gaussian part of the geodesic curvature autocorrelation function, 
and the inverse interface density.  Though
$\Gkgst$ is found to remain Gaussian on short length scales, the time
dependence of its width is no longer a $t^{1/2}$ power law, contrary to flat
systems.  For the sinusoid surface used here, it was found to be
logarithmic.  The logarithmic law is shown in the inset graph to
\fig{Gk2stRescL}, where linear regression gives $\sigma^2 \simeq (63\pm3)
\log_{10}(t/t_0)$, with $t_0=6\pm2$.  This logarithmic time dependence is
not a universal feature, as sinusoid surfaces with smaller values of
$A/\lambda$ showed power-law growth but slower than $t^{1/2}$.  It seems
likely that a {\em sub}-logarithmic growth law will be seen for larger
$A/\lambda$, similarly to results reported in \cite{oguzCTG90} on the
related Random-Field Ising Model.  The width $\sigma$ can be used as
the reference length scale $L(t)$ to rescale the geodesic curvature
autocorrelation function, as shown in
\fig{gcaf_resc_width}.  The inset shows the result for flat systems,
exhibiting dynamical scaling, within error bars.  The main graph shows the
result for the sinusoid surface.  There the curves superpose only on short
length scales, while at long length scales they systematically become wider,
with the drift being much larger than the error bars.
\begin{figure}[\figplace]
  \centerline{\psfig{figure=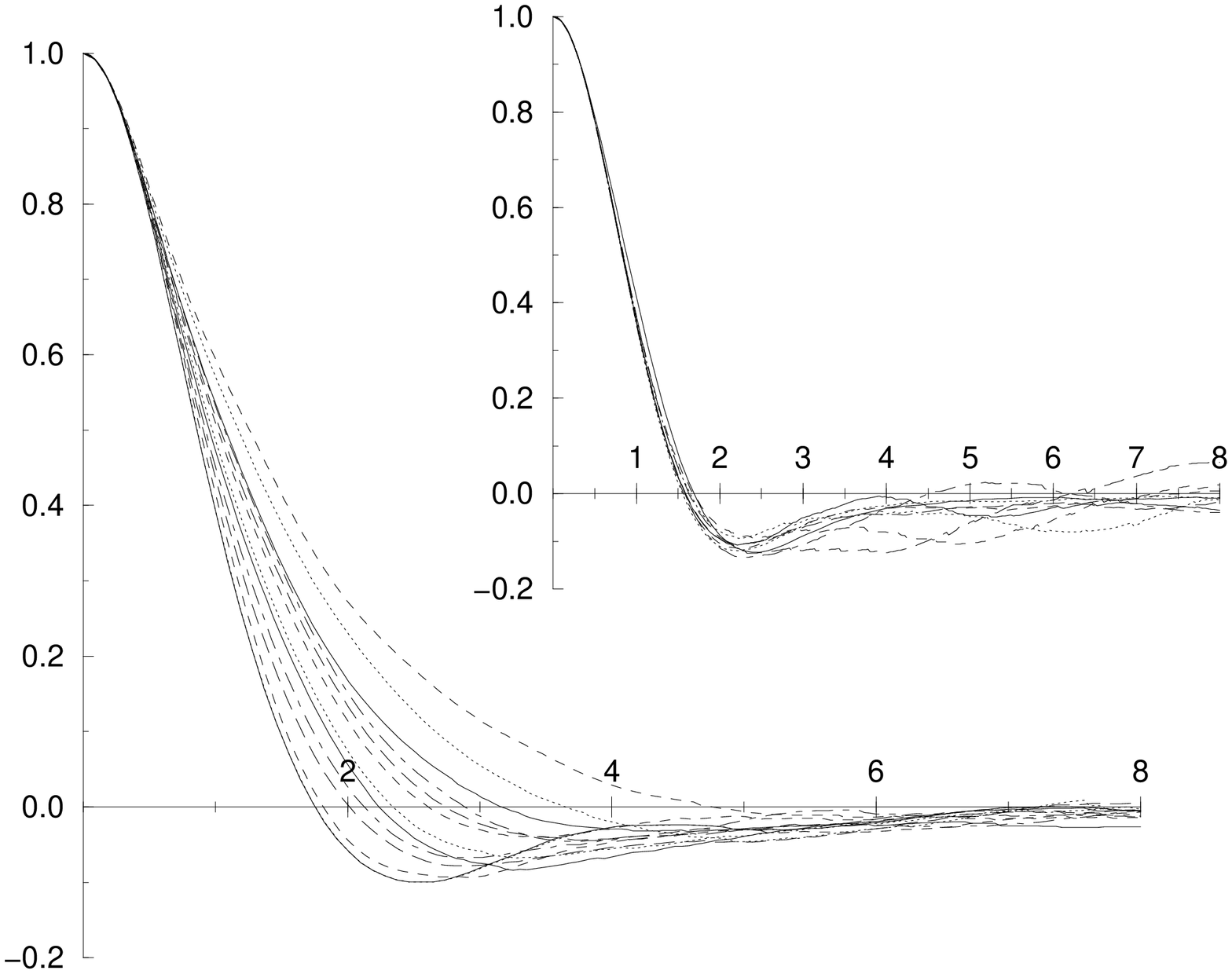,width=\figw,angle=270}}
  \caption{Main graph:  Same as main graph of \fig{Gk2stRescL} but using
	   $L(t)=\sigma(t)$ to rescale $\Gkgst$.  Inset:  Same as main graph
	   but for flat systems.}
  \label{fig:gcaf_resc_width}
\end{figure}

While the width of the geodesic curvature autocorrelation function
is a short-length-scale characteristic of the
dynamics, the interface density is more sensitive to large-scale features.
It is defined as $l(t)\equiv L(t)/A_s,$ where $L(t)$ is the total length of
interfaces at time $t$ and $A_s$ is the system area (constant in time).
Measurements of $l(t)$ for the sinusoid surface were done and compared with
those in flat systems of various sizes.  This is shown in \fig{dLdtoLofL2},
where $-d(\ln l)/dt$ is plotted as a function of $1/l^2$.
\begin{figure}[\figplace]
  \centerline{\psfig{figure=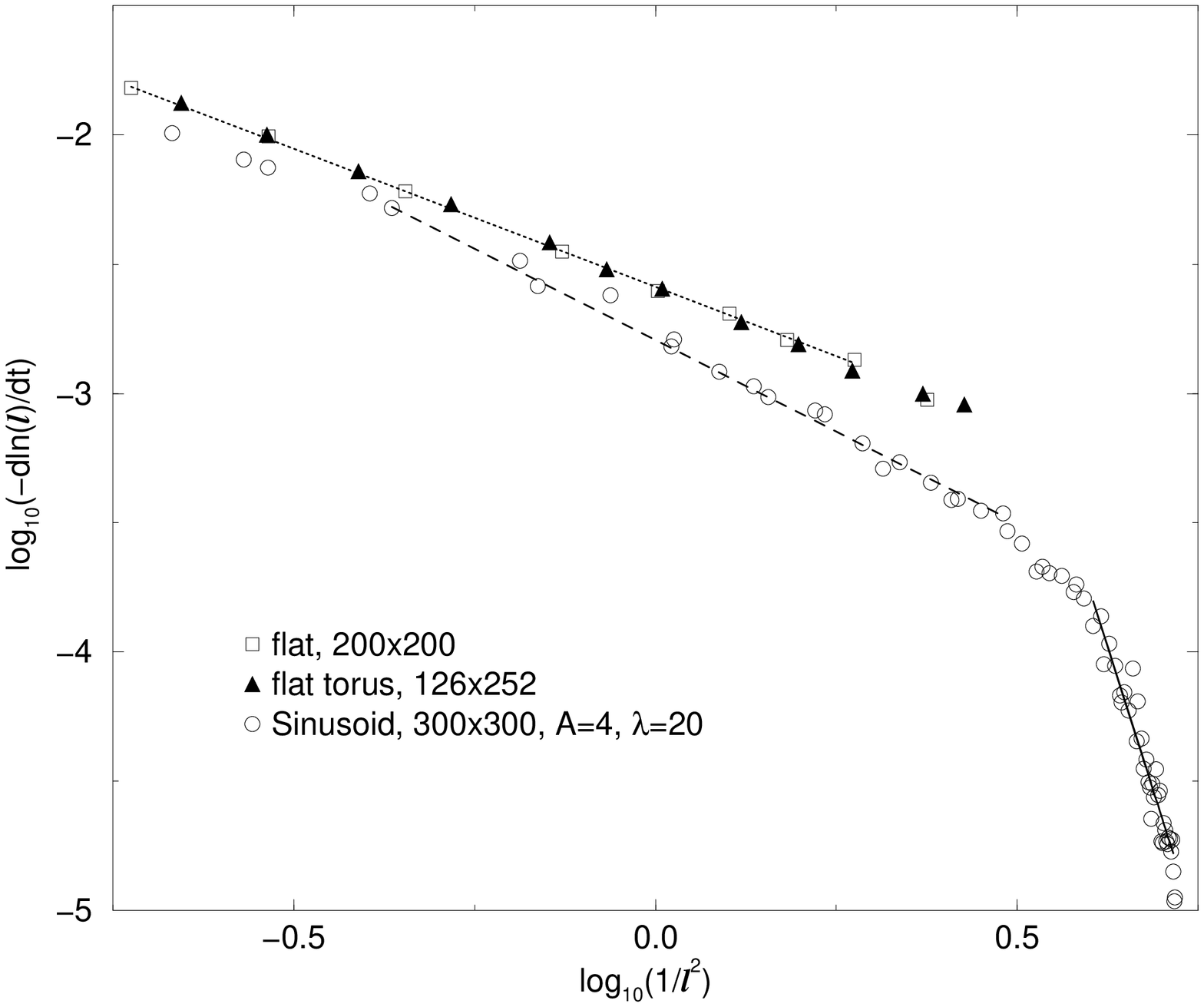,width=\figw,angle=270}}
  \caption{Logarithmic plot of $\de\ln(l)/\de t$ as a function of
	 $1/l^2$, for sinusoid surface $A=4$ and $\lambda=20$
	 (circles).  Points for flat system are triangles and squares,
	 lines are linear regressions.  Error bars (not shown) are
	 approximately same size as symbols.}
  \label{fig:dLdtoLofL2}
\end{figure}
This scale produces one unique curve for all flat systems, independent of
system size or quantity of interface.  Any straight line on this plot
indicates a power law in time for $\ell(t)$.  If the slope of the line is
denoted $m$, the intercept $b$ and $\ell_0 \equiv \ell(0)$, then
straightforward integration yields
\beq
  \ell(t) = \left(\ell_0^{2m} - 2m10^bt\right)^{1\over 2m},
\eeq
indicating $\ell \sim t^{1/2m}$ at late times.  More negative $m$ thus
corresponds to slower interface dynamics.  The interface density has units
of one over length, while the dominant length scale in Euclidean model A
grows as $t^{1/2}$.  Hence the scaling assumption valid for flat systems
predicts $\ell\sim t^{-1/2}$, i.e.~$m$ should be equal to -1 in flat
systems.  The linear regressions for the flat system curve (where bulk
integration was used rather than the interface description) gives
$m=-1.07\pm0.01$, which means a power law of $-0.46\pm0.01$ instead of -1/2.

At early times (small values of $1/\ell^2$), model-A dynamics on the
sinusoid surface also follows the same power law.  At later times however,
one can distinguish a fairly long time regime, extending from $t\simeq250$
to $t\simeq 1600$, almost a full decade of time, during which $m$ is more
negative.  The dynamics has therefore substantially slowed down.  Linear
regression in this time domain gives $m=-1.70\pm0.04$, corresponding to a
power law behavior for $\ell$ of $t^{-0.3}$.  A batch of runs for a smaller
system of size $100\times100$ gave $m=-2.0\pm0.1$, suggesting the
statistical error on the slope could be as large as 20 to 30\%.

A very interesting characteristic of this plot is the presence of a very
late time regime, where the dynamics is extremely slow, with the slope $m =
-8\pm2$.  This starts at $t\simeq 2000$, independent of system size and
discretization.  Real-time animation of moving interfaces in this regime
shows that the geodesic curvature is zero almost everywhere, but not in
sufficiently many places to completely halt the dynamics.  The interfaces,
which waver between the bumps, are very slowly hopping the bumps one by one.


\section{Metastable states}
\label{sec:metastable}

Corrugated surfaces (such as given by \eq{sinusoidk20}) allow for local
minima to appear in the configuration space of interfaces and therefore to
trap the latter in metastable states, because \eq{dLdt} implies that
Non Euclidean Model A  
interfaces can only decrease their length.  The simplest example of this
would be an ovoid interface circumventing two bumps on the sinusoid surface,
as pictured in \fig{twobumps_3D}.
\begin{figure}[\figplace]
  \centerline{\psfig{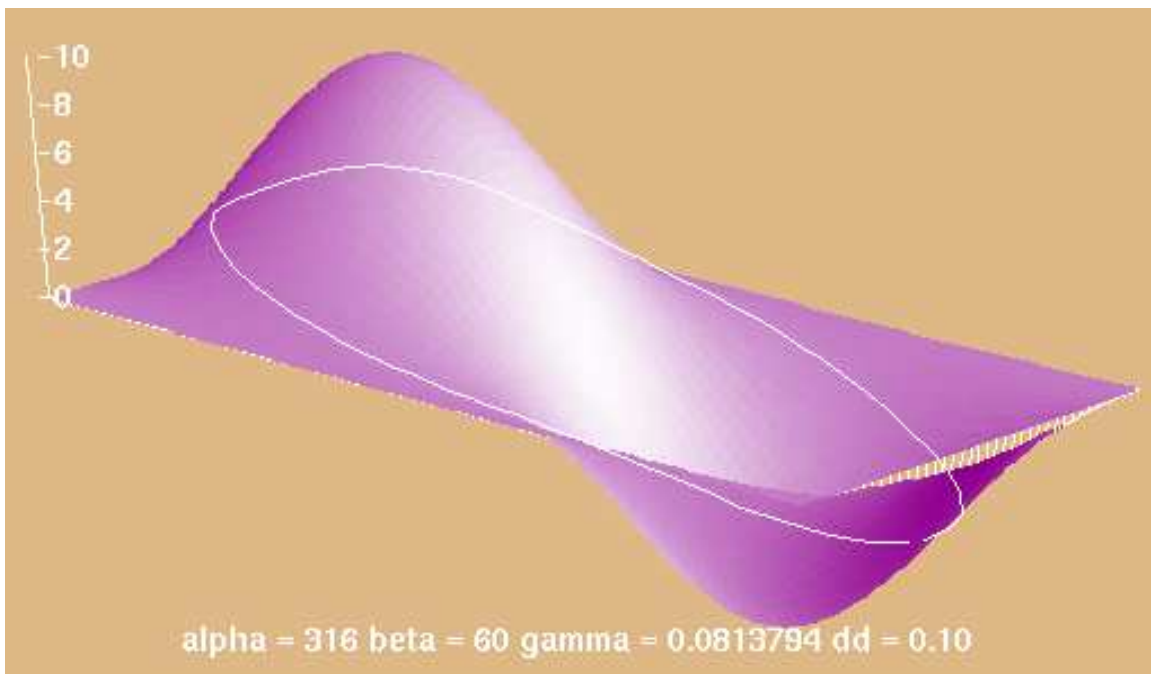}}
  \caption{Three-dimensional view of the 2-bumps surface with the ovoid
	   interface (white line).  System size $40\times20, A=10,
	   \lambda=40$.  Surface is shown as transparent to interface.}
  \label{fig:twobumps_3D}
\end{figure}
Consider the interface as it shrinks in length.  It is forced to tilt
itself, slowly moving on the outside towards the extremum of each bump.  The
total curvature at a point of the interface is the curvature as measured in
three-dimensional embedding space.  Recall that the total curvature vector
$\vK$ of a geodesic line on the surface is normal to the surface since
$K_g=0$ for a geodesic.  Therefore a necessary condition for the interface
to become stationary is for a configuration to exist where $\vK$, near both
ends of the ovoid domain, is normal to the surface.  A side view ($x-z$
projection) of the bumps with the interface is pictured in \fig{tilt2bumps}
and a top view ($x-y$ projection) in \fig{twobumps_contour}.
\begin{figure}[\figplace]
  \centerline{\psfig{figure=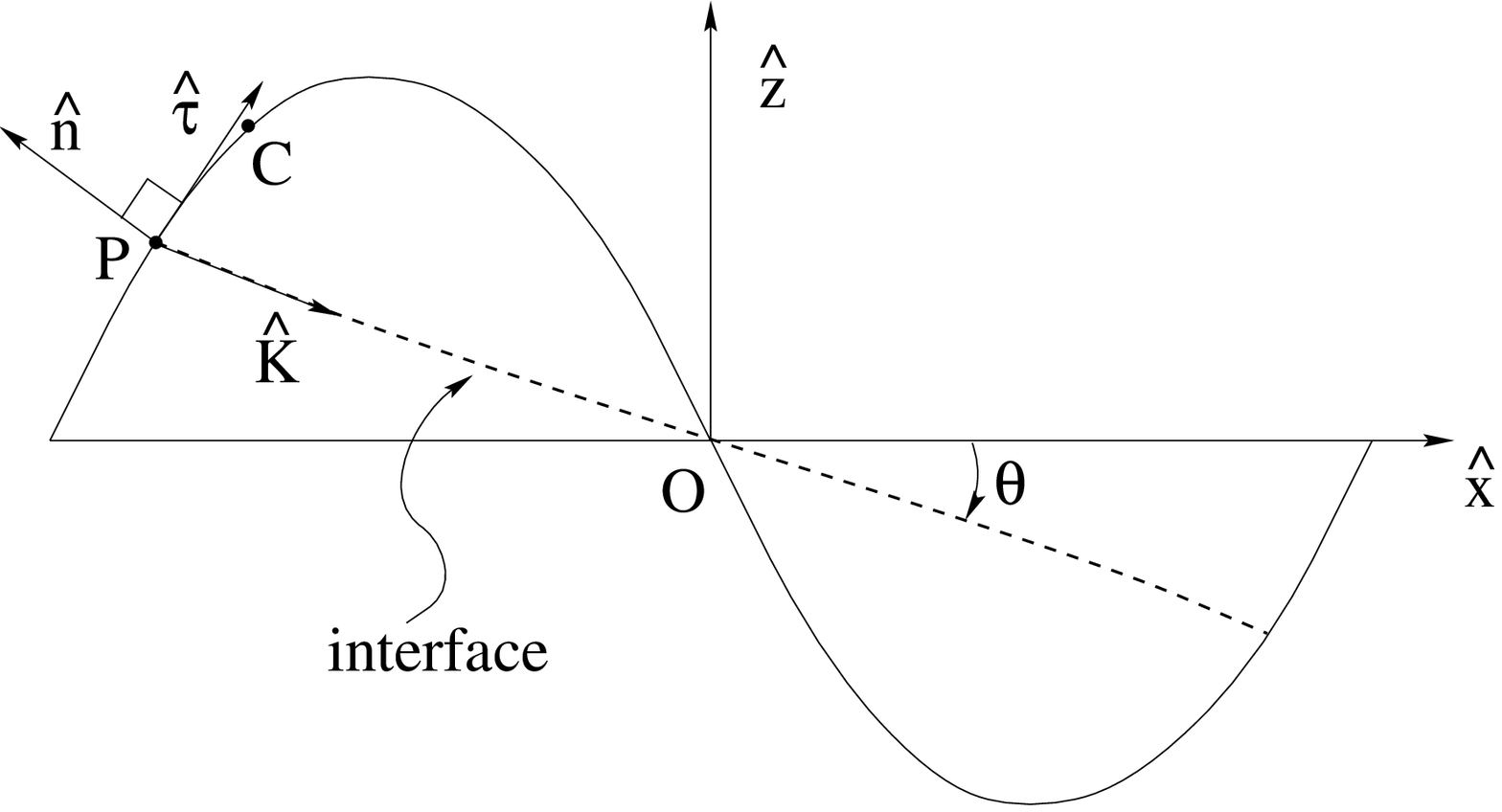,width=\figw,angle=270}}
  \caption{Side view of surface shown in \fig{twobumps_3D}
	   with projection of ovoid interface on $xz$ plane, at a later
	   time when the interface is almost stationary, i.e.~$\hat n$ and
	   $\hat K$ are almost parallel.}
  \label{fig:tilt2bumps}
\end{figure}
\begin{figure}[\figplace]
  \centerline{\psfig{figure=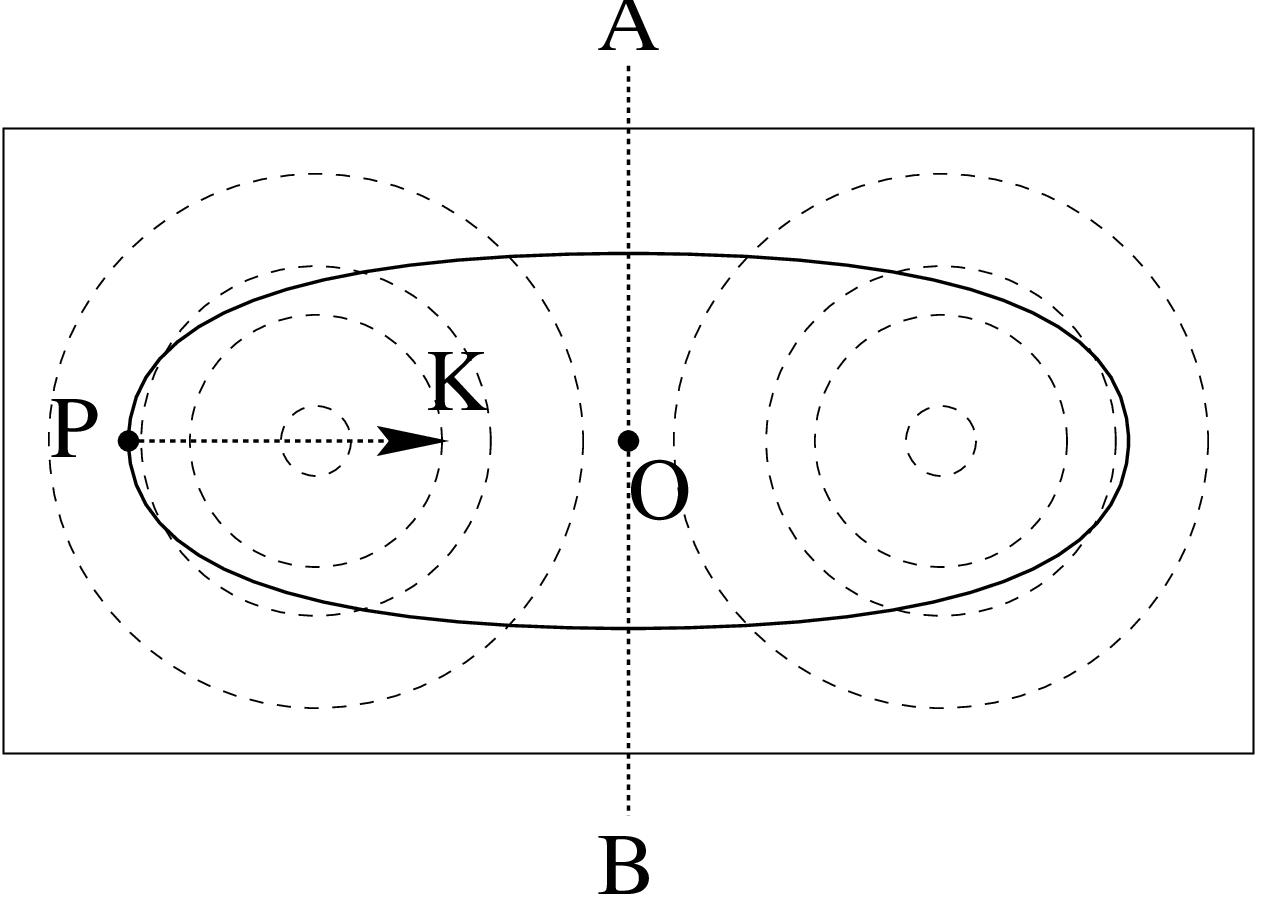,width=\figw,angle=270}}
  \caption{Top projectional view for \fig{tilt2bumps}.  Interface is 
	   thick line.  $K$ is curvature vector, other symbols correspond 
	   to those in \fig{tilt2bumps}.  Line $AB$ is ``pivot'' line 
	   for interface.  Circles are contours for the surface.}
  \label{fig:twobumps_contour}
\end{figure}
This is not a sufficient condition however since {\em all} points along the
interface must satisfy this criterion, but it does provide a minimal
condition.

We now denote the vector going from $O$ to $P$ in \fig{tilt2bumps} as $\vR
=[x,A\sin(kx)]$.  The tangent to the surface in $P$ is therefore
$\tangent=[1,Ak\cos(kx)]$, and the condition is $\vR\cdot\tangent=0$,
leading to the transcendental equation $2x=-A^2k\sin(2kx)$.  With the
substitution $x'=2kx$ and $\gamma=Ak$ this is written
\beq
  -{1\over\gamma^2} = {\sin x'\over x'}.
\eeq
This equation has a solution only when $\gamma$ is larger than a certain
value, for when $\gamma\rightarrow 0$ the 
left hand side tends to $-\infinity$ while
the right hand side is bounded between 1 and roughly $-0.22$.  The condition 
is therefore $-1/\gamma^2 \gtrsim -0.22$, or 
\beq
  {A\over\lambda} \gtrsim 0.34.
  \label{eq:hopcond}
\eeq

This was tested numerically for fixed $\lambda=40$.  Simulations give a
threshold of $0.42\pm0.02$.  This is an error of 20\%, surprisingly good
considering that the approximation uses a one-dimensional projection of the
interface on the surface:  it is not unusual for the presence of a second
dimension to give a system more freedom, thereby softening constraints such
as \eq{hopcond}.

It is important to note that \eq{hopcond} is a necessary (minimal) condition
for interfaces to get blocked around {\em two} bumps, and that blockage
around a larger number of aligned bumps occurs at similar or larger values
of $A/\lambda$.  This can be seen by considering three aligned bumps instead
of two.  The middle bump constrains the interface to belong to the $xy$
plane in its vicinity, so that point $O$ of \fig{tilt2bumps} still lies
between bumps rather than, say, at the center of the middle bump.  The
condition \eq{hopcond} therefore still holds for any number of aligned bumps
larger than or equal to two.  This is sensible since what matters is
$\lambda$, not the number of bumps.

However the situation is the reverse for bumps that are not aligned, which
is more relevant with regards to the results of the Non Euclidean Model A
simulations on
sinusoid surfaces.  Consider 4 bumps in a $2\times2$ array, and an interface
that circumnavigates the four bumps.  A top projectional view is shown in
\fig{fourbumps_contour}.
\begin{figure}[\figplace]
  \centerline{\psfig{figure=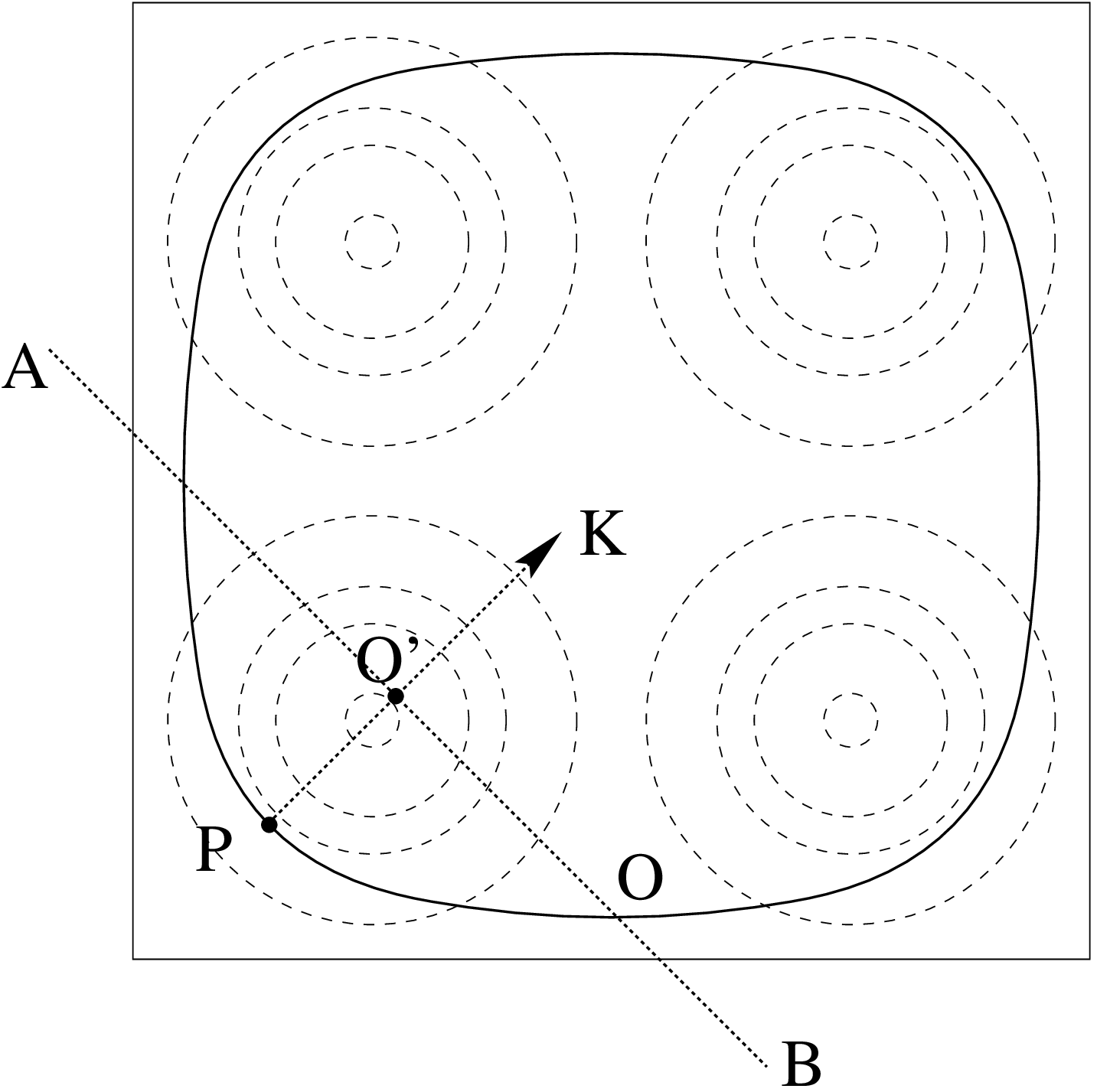,width=\figw,angle=270}}
  \caption{Top projectional view for interface around four bumps.  Same
	   conventions are used as in \fig{twobumps_contour}.  The pivot
	   point is now $O'$ rather than $O$.  }
  \label{fig:fourbumps_contour}
\end{figure}
Point $O$, one of the interface pivot points on segment $AB$, is still
halfway between bumps, near the system boundary.  The maximum-curvature
director however no longer lies along $x$ but along the line $x=y$.
Therefore, the pivot point normal to the interface at $P$ is at $O'$ rather
than $O$, decreasing the apparent wavelength by a factor of $\sqrt 2$, so
that smaller amplitudes (by the same factor) are sufficient to trap the
interface.  In this case the threshold decreases to 0.24, with simulations
giving $0.270 \pm 0.006$.  The proximity to this threshold of
$A/\lambda=0.2$ in the sinusoid simulations of the last section explains the
existence of the very-late-stage, extremely-slow regime evidenced on
\fig{dLdtoLofL2}.

The same kind of argument can be extended to, say, 6 bumps in a hexagonal
configuration, bringing $O'$ yet closer to $P$.  For a large surface
consisting of a great number of bumps, gradually increasing $A$ from zero
should cause interfaces to become stationary first around very large
conglomerates of bumps.  However the larger the conglomerate must be, the
rarer the occurrence.  As $A$ is increased further, smaller conglomerates of
bumps trap closed interfaces, until an $A$ is reached where the domains stop
growing when the dominant interface undulation length becomes comparable to
the surface $\lambda$, leading to long-range disorder.

The metastability thresholds can be generalized to more complex surfaces.
For instance, lipid bilayer membranes are usually characterized by their
bending rigidity $\kappa$.  Such membranes form random self-affine surfaces
whose average square width $W^2$ is given by\cite{peliti94}
\beq
  W^2 = {L_s^2k_BT\over 4\pi^3\kappa}
\eeq
where $L_s$ is the size of the membrane as projected on the $xy$ plane.  The
amplitude $A$ can be approximated by the width $W$ for a given size, while
the wavelength $\lambda$ can be approximated by $L_s$.  As a consequence of
\eq{hopcond}, phase ordering {\em may not occur at all} if $\kappa\lesssim
k_BT/8$, as in this case \eq{hopcond} is satisfied on {\em all}
length scales $L_s$.  As $\kappa$ is decreased towards $k_BT/8$, the 
Non Euclidean Model A 
dynamics should gradually slow down.  When $\kappa < k_BT/8$, domains might
still form but freeze in their early disordered state.

For real systems the hopping condition on $\kappa$ is not likely to be as
simple, since for such small values of $\kappa$ the surface tension $\sigma$
will usually be non negligible.  The main difference is that the threshold
involves $\kappa, \sigma$ and the length scale of interest, without
introducing any new difficulty.  This suggests that domains could order for
some time, until their dominant interface undulation length becomes larger
than a certain threshold value.

In flat systems, it was shown by Bray\cite{bray89} that model A exhibits a
zero-temperature strong-coupling fixed point.  The existence of metastable
interface configurations on corrugated surfaces implies that thermal noise
changes the quality of the dynamics once the domain interface undulations
are on the same scale as the surface corrugations, thereby eliminating the
fixed point.  Consequently, in the very-late-stage (i.e.~extremely slow)
regime observed on sinusoid surfaces, interfaces will move predominantly via
thermally-activated hopping.  Similar conclusions should be valid for model
B.  Effect of thermal noise needs to be carefully considered.

\dontput{
  Finally, the demonstration that metastable states exist allows us to
  conclude that, although in flat systems a strictly positive 
  geodesic curvature autocorrelation function 
  indicates finite-size effects dominate the dynamics, on sinusoid
  surfaces it does not.  Indeed, final system states where domains are
  much smaller than the system size and geodesically ovoid
  (i.e.~$K_g\simeq 0$ and positive everywhere) yet immobile, become
  possible.  Such states will produce a strictly positive 
  geodesic curvature autocorrelation function.  A further
  consequence of this is that the effects seen in \fig{Gk2stRescL} are
  persistent.
}

\section{Total curvature correlations}
\label{sec:totcorrel}

There is a second form of curvature autocorrelation involving the {\em
total} curvature $\vK$ of the interface, i.e.~the curvature of the interface
measured in three-dimensional space rather than in the two-dimensional space
of the manifold.  In three dimensions, a curvature {\em scalar} can not be
well defined.  The vector of curvature is the only way to properly represent
the curvature of the interface, so a dot product must be used in this case:
\beq
  \Gktotst \equiv {\langle \vK(0,t)\cdot\vK(s,t) \rangle
			  \over
		   \langle \vK(0,t)^2 \rangle}.
\eeq

Note that for {\em flat} systems, where the scalar product of two $\vK_g$ is
straightforward, {\em both} definitions of curvature autocorrelation, the
total and the geodesic one, give the {\em same} function.  It is only in
{\em curved} surfaces that they differ in very important ways.  In the
curved case, when an interface is stationary, it is so only because $K_g$ is
0 everywhere and the interface is perfectly autocorrelated, from a
two-dimensional point of view.  This perfect autocorrelation stems from the
fact that if two-dimensional observers within the manifold knew the position
and orientation of the interface at one point of the interface, they would
know the complete shape of the interface by solving the equations defining
geodesics\cite{laugwitz65} with the appropriate initial values.  The only
disorder remaining in the system is that arising from the {\em relative}
position of the domains and from the extrinsic geometry of the geodesic
lines on the surface.  This is not measurable via a geodesic curvature
autocorrelation function.

On the other hand, the {\em total} curvature of a convoluted interface at
early times, when the domains are much smaller than any geometric length
scale of the surface, is equal to the geodesic curvature of the interface,
in modulus.  Under this condition the total and geodesic curvature
autocorrelation function are nearly identical at early times (analytically
they are rigorously equal).  At late times, when an interface has become
stationary, the total curvature at a point of the interface is the curvature
of the {\em surface} along the direction of that interface.  This {\em
total} curvature autocorrelation function will therefore give information
about the curvature autocorrelations of the {\em surface} along geodesic
lines.  This information is part of the extrinsic geometry of the surface
and does not enter the geodesic curvature autocorrelation function.  The two
autocorrelation functions are therefore complementary for interface dynamics
on curved surfaces.

When the interface correlations decay faster in space than the wavelength of
the surface, $\Gktotst$ is, as discussed earlier, qualitatively very close
to $\Gkgst$.  Both correlation functions are not {\em exactly} equal as the
scalar product in $\Gktotst$ produces a slightly stronger dip amplitude by
about 20\%, and the error bars are substantially smaller in $\Gktotst$.
Near the minimum they are as much as 3 times smaller.  However, while the
average $K_g$ decreases in time, the $\vK$ eventually starts increasing
until it reaches values compatible with the surface $\lambda$.  Therefore,
$\Gktotst$ becomes very different from $\Gkgst$ at late times.

\Fig{Gk4st_k0.05} shows $\Gktotst$ for the runs on the sinusoid surface,
\eq{sinusoidk20} with $A=4$ and $\lambda=20$, at times between 17 and 10000
(17, 35, 50, 100, 150...  400, 500, 700, 1000, 1200, 1450, 1750, 2100, 2500,
3000, 4000...  10000).  The curves for $t=17$ and 10000 are labelled, with
curves at intermediate times moving gradually and monotonically from the
former to the latter.
\begin{figure}[\figplace]
  \centerline{\psfig{figure=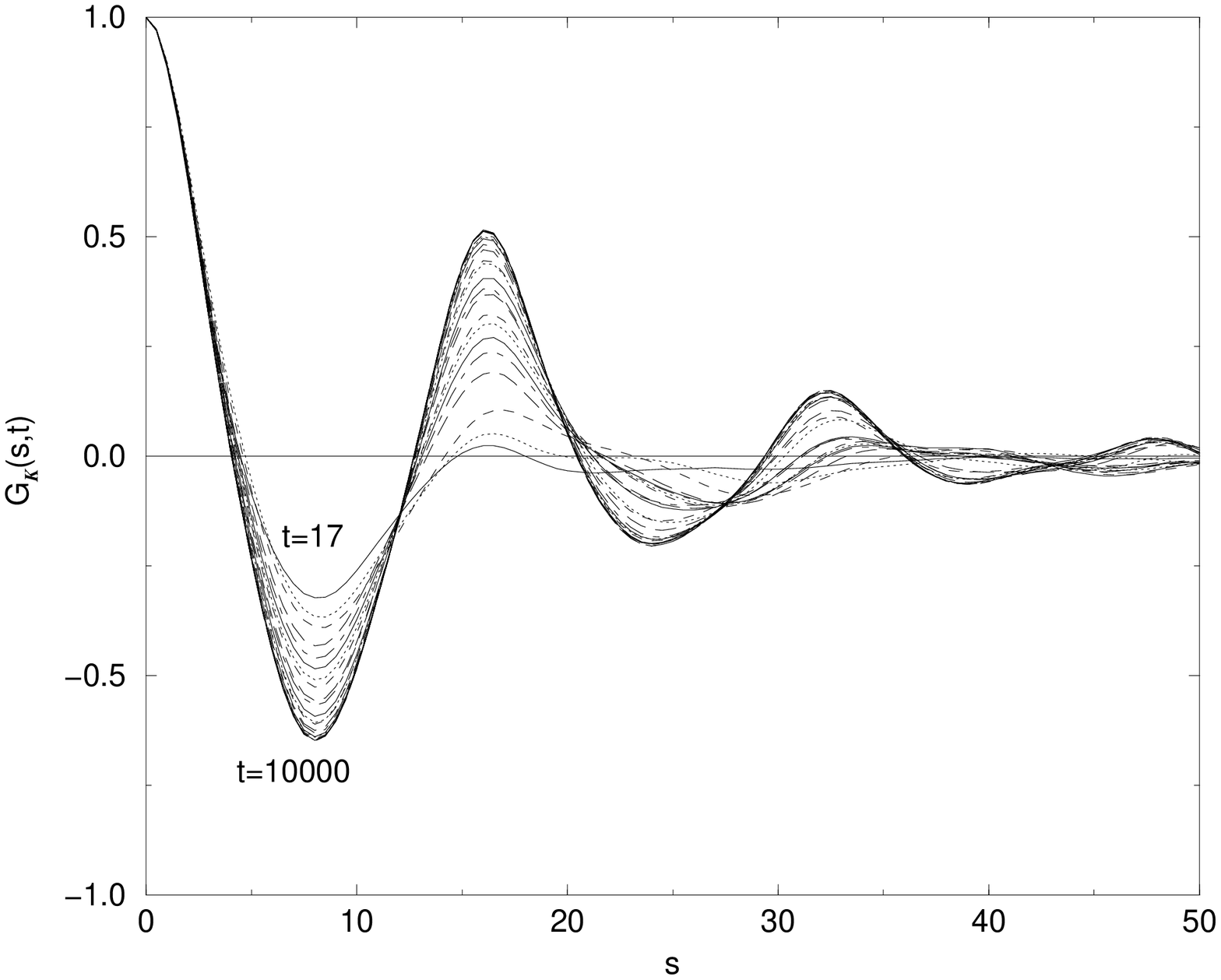,width=\figw,angle=270}}
  \caption{$\Gktotst$ for the sinusoid surface $A=4, \lambda=20$}
  \label{fig:Gk4st_k0.05}
\end{figure}
For this surface, the wavelength of the bumps is not much larger than the
initial dominant length scale of the interfaces in Euclidean systems, so
that a cross-over regime is not seen.  The correlations simply increase in
time, as more and more interfaces coarsen while being pushed between the
bumps and adopting the surface's wavelength.  At late times, the interfaces
can be approximated by a random sequence of straight lines of length
$\lambda/2$ and arcs of radius $\lambda/4$ and arclength $(\pi/4)\lambda/2$
\dontput{,as seen in \fig{waverlate}}.  This explains the peak positions in
$\Gktotst$ being roughly at integer multiples of $(\pi/4)\lambda/2$.  At
early times, only two peaks can be seen, with a third peak starting to
appear.  The emergence of the peaks at larger distances seems to coincide
with the Euclidean regime in \fig{dLdtoLofL2}.  During the slow regime
($250\lesssim t\lesssim1600$), no more peaks appear.  Those that have formed
grow slowly in amplitude and saturate.  \Fig{Gk4st_k0.025_0.02} shows
$\Gktotst$ for a sinusoid surface consisting of two wavelengths rather than
one:
\beq
  \X = [x,y,4\sin(2\pi k_1x)\sin(2\pi k_1y)+6.25\sin(2\pi k_2x)\sin(2\pi k_2y)]
  \label{eq:sinusoidk40_50}
\eeq
where $\lambda_1=2\pi/k_1=40$ and $\lambda_2=2\pi/k_2=50$.  Hence both have
$A/\lambda\simeq0.1$ and the undulations of this surface are of longer
wavelength than the previous one.
\begin{figure}[\figplace]
  \centerline{\psfig{figure=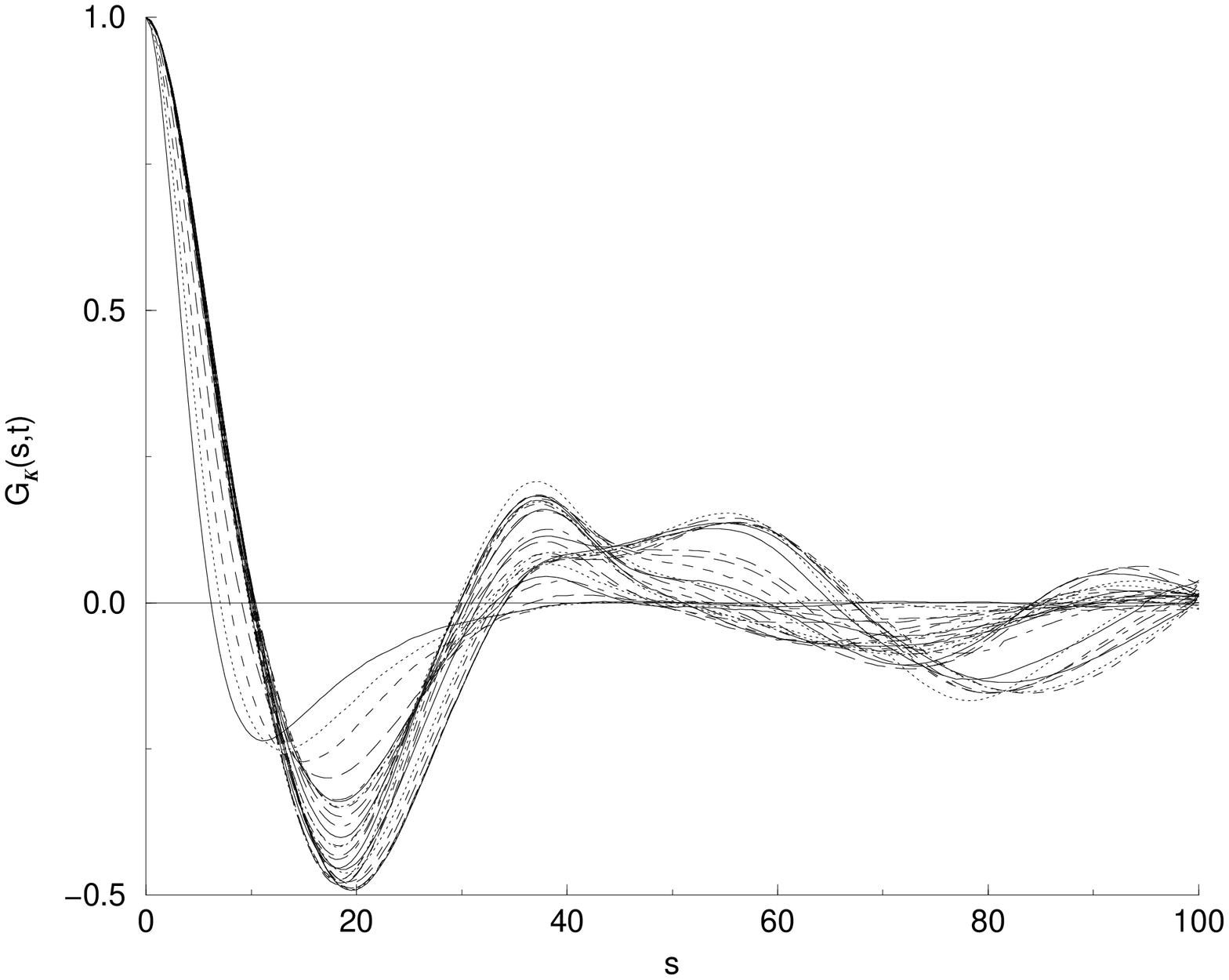,width=\figw,angle=270}}
  \caption{$\Gktotst$ for the bimodal sinusoid surface,
	   \eqprot{sinusoidk40_50} with $\lambda_1=40$ and $\lambda_2=50$.
	   Times are $t=17, 24, 35, 60, 100$ with the remaining times at
	   same values as in \fig{Gk4st_k0.05}.}
  \label{fig:Gk4st_k0.025_0.02}
\end{figure}
In this case, the earliest curve looks exactly like $\Gktwost$ in flat
systems and even has the dip positioned at roughly the same value, namely
$s\simeq10$.  Already at $t=24$ the interfaces are being pulled into the
valleys between the bumps.  This is the crossover regime.  The curvature
correlations settle into their definitive value around $t=100$, close to
$(\pi/4)\lambda/2$ with $\lambda$ around $40$.  Much longer wavelengths would be
necessary to distinctly show the three regimes:  a scaling regime at early
times during which the amplitude of the minimum would not change, then the
cross-over during which the dominant length scale saturates to a dominant
mode of the surface, and finally the late-time regime where the interfaces
move more slowly and hop the bumps.  Note that contrary to the $A=4,
\lambda=20$ case, the very-late-stage regime of \fig{dLdtoLofL2} never
occurs because $A/\lambda$ is not large enough.

%
%

\section{Summary and conclusions}

In summary, the non-Euclidean model A was used as a starting point for the
more complex, Random-Field Ising Model like\cite{hobbie96,puriCP91,oguzCTG90}
 models used in
some lipid membrane\cite{leibler86} and related surface problems.  By
deriving a set of geometric dynamical equations, using the non-Euclidean
interface velocity equation $\vv=M\xi^2\vK_g$ and developing an interface
description, numerical simulations of this difficult system could be carried
out and the results analyzed and understood qualitatively.  Moreover, the
techniques should be applicable without major impediments to dynamical
surfaces and model B on curved surfaces.

Whereas Random-Field Ising Model systems exhibit decelerated growth 
in the form of logarithmic
growth laws of various kinds, here we have shown how model A on curved
surfaces exhibit similar richness of dynamics {\em without} a bilinear
coupling to the surface:  (i) the time dependence of the amount of interface
is strongly affected by the local Gaussian curvature of the surface, (ii)
different dynamical lengths have different and even logarithmic
time-dependencies leading to the breakdown of dynamical scaling {\em
without} the disappearance of the dominant interface undulation mode, and
(iii) metastable states exist above a threshold value of $A/\lambda$,
leading to thermally activated hopping in the new very-late-stage regime.  A
more systematic study of the effect of $K_G$, a better quantitative
understanding of the observed time dependencies of $L$, and experimental
methods to obtain the dominant interface undulation mode, would be useful at
this point.

\section*{Acknowledgments}

The authors wish to thank Dr.~Mohamed Laradji and Fran\c{c}ois L\'eonard for
helpful discussions, Prof.~C.~M.~Knobler for useful comments, and the
Natural Sciences and Engineering Research Council of Canada (NSERC), the
Walter C.~Sumner Foundation and the Fonds pour la Formation de Chercheurs et
l'Aide \`a la Recherche (FCAR) for partial funding of parts of this work.

%
%

\smallersize

\begin{thebibliography}{10}
\bibitem{defineKG}
See references \cite{laugwitz65} and \cite{david88} for further
discussion of these notions.  The total interface curvature $\vK$ is the
curvature at a point of a line interface in three-dimensional space.  Its
geodesic curvature at that point is that component of the total curvature
which lies within the surface.  On the other hand, a surface has two
orthogonal radii of curvature at every point on it, which can be of the same
or opposite sign.  The Gauss curvature $K_G$ is the product of these two
curvatures.  $K_G$ vanishes for a flat region of a surface, is positive in
regions where the surface looks locally spherical, and is negative if the
region is locally saddle-like.
\bibitem{errbars}
The vertical error bar is 0 at $s=0$ and increases roughly linearly to a
little under 0.02 at the minimum of $\Gkgst$.
\end{thebibliography}

{}{

}

\end{document}